%% file: main.tex
\documentclass[
    twocolumn,
	prd,
    amssymb,
	preprintnumbers,
	secnumarabic,
	nofootinbib,
	superscriptaddress]{revtex4-1}

\pdfoutput=1

\usepackage{graphicx}
\usepackage{enumitem}
\usepackage{latexsym}
\usepackage{amsfonts}
\usepackage{amssymb}
\usepackage{array}
\usepackage{pifont}
\usepackage{color}
\usepackage{xcolor}
\usepackage{amsmath}
\usepackage{slashed}
\usepackage{dcolumn}
\usepackage{verbatim}
\usepackage{float}
\usepackage{multirow}
\usepackage{xspace}
\usepackage[normalem]{ulem}
\usepackage{hyperref}
\usepackage{subfigure}
\usepackage{anyfontsize}
\usepackage{t1enc}

\definecolor{aquamarine}{rgb}{0.2,0.7,0.6}
\definecolor{mycerulean}{RGB}{0,166,214} 
\definecolor{hypershade}{rgb}{0.3,0.3,0.8}
\definecolor{subtlered}{rgb}{0.8,0.3,0.3}

\hypersetup{
  pdfauthor={Nirmal Raj},
  pdftitle={$\nu$isance},
  pdfsubject={Nuisance},
  colorlinks=true,
  citecolor=aquamarine,
  urlcolor=mycerulean,
  linkcolor=gray
}

\input{universalnewcommands.tex}

\def\rhox{\rho_\chi}
\def\mx{m_\chi}
\def\mdm{m_\chi}

\def\Mdet{M_{\rm fid}}
\def\texp{t_{\rm exp}}
\def\Rdet{R_{\rm fid}}
\def\Erec{E_{\rm R}}
\def\nT{n_{\rm T}}
\def\rhoT{\rho_{\rm T}}
\def\mT{m_{\rm T}}
\def\muTx{\mu_{\rm T\chi}}
\def\muNx{\mu_{\rm N\chi}}

\def\sigmaTx{\sigma_{\rm T\chi}}
\def\sigmaNx{\sigma_{\rm N\chi}}

\def\sigmaTxeff{\sigma_{\rm T\chi}^{\rm eff}}

\def\Nhit{N^{\rm exp}_{\rm hit}}

\def\NBG{N_{\rm B}}

\def\NexpCL{N_{\rm exp}^{\rm 90CL}}

\allowdisplaybreaks

\begin{document}

\title{The neutrino roof: Single-scatter ceilings in dark matter direct detection}

\author{Nirmal Raj}
\email{nraj@iisc.ac.in}

\author{Biprajit Mondal}
\email{biprajitm@iisc.ac.in}

\affiliation{Centre for High Energy Physics, Indian Institute of Science, C. V. Raman Avenue, Bengaluru 560012, India}

\date{\today}

\begin{abstract}
We identify the maximum cross sections probed by single-scatter ``WIMP'' searches in dark matter direct detection.
Due to Poisson fluctuations in scatter multiplicity, these  ceilings scale logarithmically with mass for heavy dark matter and often lie in regions probed by multiscatter searches.
Using a generalized formula for single-scatter event rates we recast WIMP searches by the quintal-to-tonne scale detectors 
XENON1T, XENONnT, LZ, PANDAX-II, PANDAX-4T, DarkSide-50 and DEAP-3600 to obtain ceilings and floors up to a few $10^{17}$~GeV mass and $10^{-22}$~cm$^2$ per-nucleus cross section.
We do this for coherent, geometric, isospin-violating xenophobic and argophobic spin-independent scattering, and neutron-only and proton-only spin-dependent scattering.
Future large-exposure detectors would register an almost irreducible background of atmospheric neutrinos that would determine a dark matter sensitivity ceiling that we call the “neutrino roof”, in analogy with the well-studied “neutrino floor”.
Accounting for this background, we estimate the reaches of the 10$-$100 tonne scale DarkSide-20k, DARWIN/XLZD, PANDAX-xT, and Argo, which would probe many decades of unconstrained parameter space up to the Planck mass, as well as of $10^3-10^4$ tonne scale noble liquid detectors that have been proposed in synergy with neutrino experiments.
\end{abstract}

\maketitle

\section{Introduction}

In full vigor goes the hunt for the elusive particle species making up dark matter.
Underground and terrestrial surface laboratories typically search for dark matter (DM) that scatters at most once per transit, as that is the signature of the well-studied weakly-interacting massive particles (WIMPs).
Though WIMP masses are thought to reside around the electroweak scale (GeV$-$TeV), it was recognized even in the very first proposal of direct searches~\cite{GoodmanWitten:1984dc} that these detectors admit enough flux to reach masses $\gsim10^{13}$~GeV.
Candidates for such DM include 
WIMPzilla-like~\cite{Models:Chung1998wimpzillas,*Models:kolb1998wimpzillas,*Models:Harigaya2016:GUTzillas}, colored~\cite{Models:ColoredDM},
and 
baryon-charged~\cite{Models:BNL:DarkBaryonGeVMediator} states, 
composite nuclei~\cite{Models:nucleiHardy:2014mqa,*Models:nucleiHardy:2015boa,*Models:nucleiMonroe:2016hic,*Models:Nuggets},
dark monopoles~\cite{Models:EWSymMonopoles:Bai:2020ttp}, electroweak-symmetric solitons~\cite{Models:ElectroweakBalls}, Planck-scale black hole relics~\cite{Models:PlanckScaleBHRelicsBaiOrlofsky2019,*Models:PlanckScaleBHRelicsSantaCruz}, 
and even elementary DM that interacts with nuclei via effective operators~\cite{Bramante:2018tos};
these may be formed by a number of cosmological mechanisms~\cite{snowmass:Carney:2022gse}.
In the wake of multiple null results in searches for weak-scale DM and the concomitant rise in prominence of ultra-heavy DM, experimenters have begun to aim for very high masses. 
As the sensitivities to scattering cross sections weaken with increasing DM mass, these searches naturally tend to cross sections large enough for multiple scatters per transit to occur. 
Accordingly, dedicated multiscatter searches in recent times have placed limits on DM-nucleus cross sections $\gsim \Oc$(barn) for up to $10^{19}$~GeV mass scales~\cite{DAMA1999,EdelweissCDMSAlbuquerqueBaudis2003,KavanaghSuperheavy:2017cru,PICO:MS:Broerman2022,XENON1T:MSSSprojexn:Clark:2020mna,PlasticEtch:Bhoonah2020fys,CollarBeacomCappiello2021, DEAP:MS:2021raj,XENON1T:MS:2023iku,LZ:MS:2024psa,snowmass:Carney:2022gse}. 

In this work we derive the full extent of {\em single}-scatter constraints of ultra-heavy DM by recasting existing WIMP searches, as well as the future reaches of $\Oc(10-100)$ tonne detectors such as DARWIN/XLZD~\cite{DARWIN:Macolino:2020uqq,XLZD:Baudis:2024jnk}, PANDAX-xT~\cite{PandaX-xT:2024oxq},
DarkSide-20k~\cite{DarkSide-20k:2017zyg},
and ARGO~\cite{ARGO:2018,*ARGOSnowmassLOI}.
The key feature of our result is the presence of ceilings in cross section vs DM mass space, above which single-scatter searches are insensitive.
Beyond these ceilings, searches for multiscatter signatures are required to uncover DM interactions.
Thanks to Poisson fluctuations in the number of DM scatters per transit (i.e., in ``multiplicity''), single-scatter and multiscatter sensitivity regions typically overlap.
As it is these Poisson fluctuations that set the single-scatter ceiling, we find that it scales {\em logarithmically} with DM mass.
In contrast, due to the DM flux falling inversely with the DM mass,
the usually quoted upper bounds on the cross section scale {\em linearly} with DM mass.
Single-scatter ceilings were shown in the experimental results of Refs.~\cite{XENON1T:MSSSprojexn:Clark:2020mna,XENON1T:MS:2023iku,LZ:MS:2024psa}; in this work, we will provide a simplified, general prescription to obtain these ceilings at any DM direct detection experiment, and use it to estimate constraints at recent and forthcoming noble-liquid detectors.
These are significant improvements over the mass-independent single-scatter ceilings shown in Ref.~\cite{Bramante:2018qbc}.

Single-scatter ceilings naturally lend themselves to the idea of a ``neutrino roof'', in analogy with the neutrino floor.
The neutrino floor is defined variously~\cite{nufloor:Strigari:2009bq,nufloor:Billard:2013qya,nufloor:OHare:2021utq}, but the general idea is that it is a curve on the DM cross section vs mass plane that represents the discovery reach of DM in the presence of an irreducible background of neutrinos sourced by the Sun, terrestrial atmosphere-cosmic ray interactions, and cosmic supernovae.
As this curve is simply a mapping between a particular DM recoil energy spectrum and the cross section vs mass plane, and since for every DM mass there are two values of the cross section (the upper and lower bounds) that will generate identical recoil energy spectra, the map should have both a floor and a roof.
We will not concern ourselves with the exact definition and location of the neutrino roof, unlike Refs.~\cite{nufloor:Billard:2013qya,nufloor:OHare:2020lva,nufloor:OHare:2021utq}
that charted the neutrino floor by varying the detector exposure.
Instead, we will assume representative values of detector exposures that can be realistically achieved and estimate the DM sensitivity ceilings accounting for atmospheric neutrino backgrounds with their statistical and systematic uncertainties.
These exposures are about 3$-$50 times that of the 3rd generation (Gen-3) DM detectors DARWIN/XLZD, PANDAX-xT and ARGO; the feasibility of such detectors, e.g., acquisition of the required quantities of noble elements, readout technologies, etc., are being actively researched in synergy with neutrinoless double-beta decay ($0\nu\beta\beta$) proposals~\cite{Xekton:Avasthi:2021lgy,*Xekton:Anker:2024xfz} and the DUNE experiment~\cite{DUNE:2020lwj,DUNEModuleDM:PNL2020,*DUNEModuleDM:snowmass:Avasthi2022,*DUNEModuleDM:Bezerra2023}.

Single-scatter ceilings and/or neutrino roofs can be crucial diagnostic devices.
If future detectors register DM signals significantly above neutrino backgrounds, a statement can be made one way or another about the DM cross section and mass by immediately following up with multiscatter searches.
This is because in such searches measuring the multiplicity points to the scattering cross section, and the event count points to the DM mass~\cite{Bramante:2018qbc}.
If no multiscatter signal is registered, one may be sure that the single-scatter signal came from a region close to the neutrino floor as opposed to the roof.
Of course, null results in single-scatter searches can also be verified with multiscatter searches, and the order in which these two classes of searches are performed can also be reversed.

To illustrate the general nature of our prescription, we derive single-scatter direct detection limits on ultra-heavy DM for multiple scenarios: scattering that is spin-independent with and without nuclear-coherent enhancement, isospin-violating with focus on Xe-phobic and Ar-phobic cases, and spin-dependent with focus on neutron-only and proton-only scattering as reported by experiments.
These cases help us compare results with the literature and between xenon and argon targets, but in general other scalings between nucleon and nuclear cross sections are possible, so it is important to interpret limits on individual nuclear targets independently.

This paper is laid out as follows.
In Section~\ref{sec:basics} we provide our master formula for estimating the full extent of single-scatter constraints, and discuss the scaling of the single-scatter ceiling cross section with various quantities of interest.
In Section~\ref{sec:results} we describe limits from current and upcoming experiments, and show projections for even larger exposures.
In Section~\ref{sec:disc} we provide discussion on the scope of our work, and conclude.
In the appendixes we illustrate the unique dependence of single-scatter ceilings on detector mass as opposed to exposure, and review single-scatter and multiscatter search techniques used by noble liquid experiments. 

\section{Set-up}
\label{sec:basics}

\subsection{Cross sections, multiplicities, event rates}
\label{subsec:masterformulae}

\begin{table*}[]
    \centering
    \begin{tabular}{|p{34mm}|p{31mm}|p{18mm}|p{14mm}|p{5mm}|p{40mm}|}
    \hline
   &  & $\Mdet \times \texp$  & &  &  \\
  target  & detector  & (ton $\times$ yr) & $\Erec$ (keV) & $\epsilon_{\rm NR}$ &  ($N_{\rm B}\pm\sigma_B$, $N_{\rm obs}, N_{\rm exp}^{\rm 90CL})$  \\
     \hline
       &  {\bf DARWIN/XLZD}~\cite{DARWIN:Macolino:2020uqq} & 40 $\times$ 5 & [5, 35] & 0.50 & (4.1, 4.1, 4.0) \\
       & {\bf PANDAX-xT}~\cite{PandaX-xT:2024oxq} & 34.2 $\times$ 5.85 & [4, 35] & 0.50 & (48$\pm$6.9, 48, 11.4) \\
    xenon   & XENON1T~\cite{XENON1T:MS:2023iku} & 1.3 $\times$ 0.76 & [10, 40] & 0.80 & (7.4$\pm$0.6, 14, 12.8) \\
  \scriptsize{2.94 g/cm$^3$}   & XENONnT~\cite{XENONnT:SS:2023cxc} & 4.18 $\times$ 0.26 & [10, 40] & 0.80 & (2.03$\pm$0.16, 3, 4.7) \\
  \scriptsize{$^{128}$1.9\%, $^{129}_{\rm SD}26.4\%$, $^{130}$4.1\%}  & LZ~\cite{LZ:SS:2022lsv} &  5.5 $\times$ 0.16 & [5, 50] & 0.90 & ($-$,$-$, 4.4)\\
   \scriptsize{ $^{131}_{\rm SD}$21.2\%, $^{132}$26.9\%,}    & PANDAX-II~\cite{PandaX-II:SS:2020oim} &0.33 $\times$ 1.1 & [10, 30] & 0.85 & (40.3$\pm$3.1, 38, 7.8)\\
   \scriptsize{ $^{134}$10.4\%, $^{136}$8.9\%}    & PANDAX-4T~\cite{PandaX-4T:SS:2021bab} & 2.67 $\times$ 0.24 & [30, 90] & 0.75 & (9.8$\pm$0.6, 6, 0.8)\\
    & {\fontsize{3}{3.6} \selectfont KILOXENON} & 40 $\times$ 25 & [5, 35] & 0.50 & (20.6$\pm$6.1,20.6$\substack{\times2 \\ \div 2}$,$10.9\substack{+6.3 \\-6.7}$)\\
   & \tiny{MYRIAXENON} & 10$^3$ $\times$ 10 & [5, 35] & 0.50 & (206$\pm$44,206$\substack{\times2 \\ \div 2}$,$60\substack{+44\\-45}$)\\
        \hline
          & {\bf DarkSide-20k}~\cite{DarkSide-20k:2017zyg} & 20 $\times$ 10 & [30, 200] & 0.90 & (3.2, 3.2, 3.7) \\
     & {\bf Argo}~\cite{ARGO:2018,*ARGOSnowmassLOI} & 300 $\times$ 10 & [55, 100] & 0.90 & (15.6$\pm$5, 15.6, 6.4) \\
    argon     & DarkSide-50~\cite{DarkSide:SS:2018kuk} & 0.031 $\times$ 1.46 & [80, 200] & 0.70 & (0, 0, 2.3) \\
   \scriptsize{1.40 g/cm$^3$}      & DEAP-3600~\cite{DEAP:SS:2019yzn} &0.824 $\times$ 0.63 & [70, 100] & 0.24 & (0, 0, 2.3) \\ 
        \scriptsize{$^{36}$0.33\%, $^{38}$0.06\%, $^{40}$99.6\%}   &  \tiny{MYRIARGON} & 300 $\times$ 33.3 & [55, 100] & 0.90 & (51.5$\pm$12.6,51.5$\substack{\times2 \\ \div 2}$,$19.7\substack{+13.1 \\-13.2}$)\\
             & \tiny{DECIMEGARGON} & 10$^4$ $\times$ 10 & [55, 100] & 0.90 & (515$\pm$106,515$\substack{\times2 \\ \div 2}$, 174$\substack{+106\\-106}$)\\
        \hline
    \end{tabular}\\ \vspace{.5cm}
    \includegraphics[width=\textwidth]{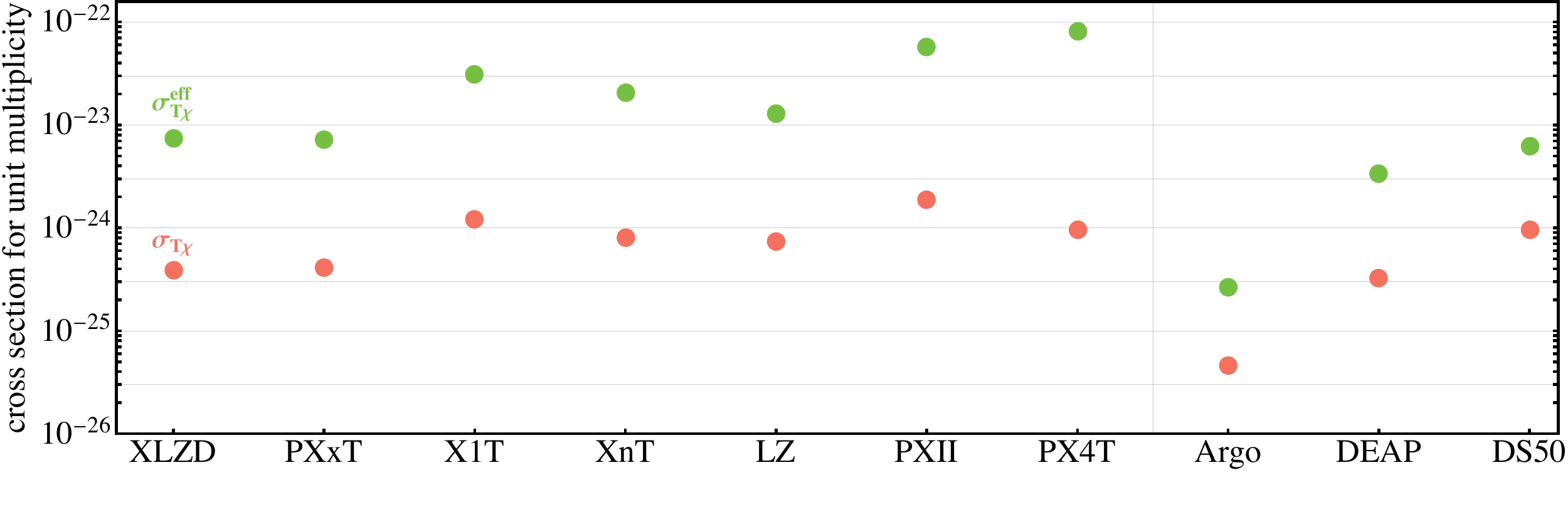}
    \caption{{\bf \em Top.}
     Experiments (with proposed forthcoming ones in {\bf bold}) for which single-scatter upper and lower limits are derived and shown in Figs.~\ref{fig:limits} and \ref{fig:nuroof}, along with their detector fiducial mass $\Mdet$, live time $\texp$, signal efficiency/nuclear recoil acceptance $\epsilon_{\rm NR}$, background estimate $\NBG$ with uncertainty $\sigma_B$, observed event count $N_{\rm obs}$, and the 90\% C.L. limit $\NexpCL$ as obtained from Eq.~\eqref{eq:def:90CL}.
      For DARWIN, DarkSide-50, DEAP-3600 and DarkSide-20k, the background uncertainty is negligible and Eq.~\eqref{eq:def90CL:noBGerror} is used instead.
      For LZ we directly take the statement in Ref.~\cite{LZ:MS:2024psa} that $N_{\rm obs} = 4.4$.
      PANDAX-xT quotes its $\Erec$ range as 1$-$10 keV electron equivalent, which translates to 4$-$35 keV nuclear recoils~\cite{BaudisTalks,DARWINish:BaudisSchumann:2015cpa}.
      In the first column noble liquid densities and isotope abundances are listed; xenon isotopes that have spin-dependent interactions are marked with the subscript ``SD''. 
     See Sec.~\ref{sec:results} for further details and Appendix~\ref{app:searchmethods} for these experiments' search techniques.
      {\bf \em Bottom.} The effective cross sections in Eq.~\eqref{eq:sigmaTxeff} for spin-independent scattering that give unit multiplicity in various detectors with the specifics given in the top panel, and analogous total DM-nucleus cross sections. 
      The finite range of nuclear recoils detectable and form factor suppression render $\sigmaTxeff$ smaller than $\sigmaTx$ by an order of magnitude.}
    \label{tab:detectordetails}
\end{table*}

We now set up our central formalism, which generalizes the treatment of WIMP direct detection commonly seen in the literature.  
The total number of events of DM undergoing single scatters within a recoil energy range $\{ E_{\rm R,min}, E_{\rm R,max}\}$ on a nuclear target $T$ in a detector fiducial volume of radius $R_{\rm fid}$ over a live time $t_{\rm exp}$ is
\beq
N_{\rm ev}^{\rm SS} = \epsilon_{\rm NR} \  p_{\rm hit}(1, \Nhit) \  \Phi_{\rm int}~,
\label{eq:NevSSmaster}
\eeq
where $\epsilon_{\rm NR}$ is the detector efficiency in the $\Erec$ range,
\beq
\Phi_{\rm int} = \bigg(\frac{\rhox}{\mx}\bigg) \pi \Rdet^2 \bar v \texp
\label{eq:integfux}
\eeq
is the integrated flux of DM particles of density $\rhox = 0.3~$GeV/cm$^3$ and mass $\mx$ through the detector, with the average DM speed $\bar v = 270$~km/s.
We assume spherical detectors, which gives results in excellent agreement with those reported by experiments; detailed treatments of detector geometries are best left to search collaborations.
Hence the detector fiducial mass is given by $\Mdet = 4\pi \rhoT \Rdet^3/3$ for target density $\rhoT$. 
The multiplicity, {\em i.e.} the average number of DM-nucleus scatters per transit in the recoil energy range, is
\beq
\Nhit = \sigmaTxeff \nT L_{\rm ave}~,
\label{eq:multiplicity}
\eeq
and the Poisson probability that $k$ scatters are obtained when $\lambda$ scatters are expected is
\beq
p_{\rm hit} (k, \lambda) = \frac{\lambda^k e^{-\lambda }}{k!}~. 
\label{eq:probmultiplic}
\eeq
Here $\nT$ is the nuclear target number density and $L_{\rm ave} = 4 \Rdet/3$ is the average detector chord length.  
The effective cross section
\beq
\sigmaTxeff = \sigmaTx \frac{\mT}{2\muTx^2 \bar v} \int^{E_{\rm R, max}}_{E_{\rm R, min}} d\Erec S(\Erec) \eta(\Erec)~,
\label{eq:sigmaTxeff}
\eeq
with 
$\muTx$ the DM-target reduced mass,
and $\eta$ the usual velocity integral given by 
\bea
&& \eta (\Erec) = \int_{v_{\rm min}(\Erec)}^{v_{\rm esc}}  d^3 v \frac{f_{\rm lab} (\Vec{v})}{v}\\
 &= & \nn \frac{\mathcal{N}}{v_0} \begin{cases}
0 \ \ \ \ \ \ \ \ \ \ \ \ \ \ \ \ \ \  \ \ \ \ \ \ \ \  , v_{\rm min} > v_{\rm esc} + v_{\rm E},\\
\frac{{\rm erf}(z_{\rm min}) - {\rm erf}(z_{\rm min}-z_{\rm E})}{2z_{\rm E}} \\ 
- \frac{e^{-z_{\rm esc}^2}}{\sqrt{\pi}} [\frac{z_{\rm esc} - z_{\rm min}}{z_{\rm E}}+1], - v_{\rm E} < v_{\rm min} -  v_{\rm esc} < v_{\rm E},\\
\frac{{\rm erf}(z_{\rm min}+z_{\rm E}) + {\rm erf}(z_{\rm min}-z_{\rm E})}{2z_{\rm E}} \\ - \frac{2}{\sqrt{\pi}}e^{-z_{\rm esc}^2} \ \ \ \ \  \ \ \ \  \ \ \  \ \  , v_{\rm min} < v_{\rm esc} - v_{\rm E}, \\
\end{cases}
\eea
where in the second line we have expressed the analytic form for a truncated Maxwell-Boltzmann distribution in the Earth's reference frame~\cite{nufloor:GaspertGiampaMorrissey:2021gyj}, and $v_{\rm min} (\Erec) = \sqrt{\mT \Erec/2\muTx^2}$.
Here $z_i \equiv v_i/v_0$, 
$\mathcal{N}^{-1} = {\rm erf}(z_{\rm esc}) - 2 z_{\rm esc} e^{-z^2_{\rm esc}}/\sqrt{\pi}$,
$v_{\rm esc} = 544$~km/s is the Galactic escape speed,
$v_{\rm E} = 254$~km/s is the net Earth speed relative to the DM halo, and
$v_0 = 238$~km/s is the local standard of rest at the solar position~\cite{Baxter:2021pqo}.
For spin-independent scattering, $S(\Erec) \to F^2(\Erec)$ with $F$ the Helm form factor~\cite{Helm:1956zz,Lewin:1995rx}.
For spin-dependent scattering, $S(\Erec)$ is the structure factor, which we adopt from the parametric fits in Ref.~\cite{spindepstrucfac:Klos:2013rwa} by making the substitution $u \to \mT \Erec b^2$.
To connect with how results are displayed by experiments, we will consider ``neutron-only'' and ``proton-only'' structure factors.

In contrast to the total DM-nucleus cross section $\sigmaTx$, the effective cross section $\sigmaTxeff$ determines the multiplicity as seen by the detector within a finite range of recoil energies over which a search is performed, and accounts for form factor suppression integrated over the DM velocity distribution.
Thus $\sigmaTxeff < \sigmaTx$, so that the multiplicity is smaller than one would naively expect from the simpler and more widely used $\sigmaTx$.
We show this effect in the bottom panel of Table~\ref{tab:detectordetails} by marking the spin-independent $\sigmaTxeff$ for which $\Nhit = 1$ in Eq.~\eqref{eq:multiplicity} for the various experiments listed in the top panel and using the $\Erec$ range specified there, as well as the cross section obtained if we had replaced $\sigmaTxeff \to \sigmaTx$ in Eq.~\eqref{eq:multiplicity}.
This makes for an important difference between the results of this work and the estimates of Ref.~\cite{Bramante:2018qbc}.

In the limit of small cross sections where the detector is optically thin to DM ($\Nhit \ll 1$), we have $p_{\rm hit} (1,\Nhit) \simeq \Nhit$ per Eq.~\eqref{eq:probmultiplic}, and our Eq.~\eqref{eq:NevSSmaster}, when expressed as a differential rate per target nucleus, reduces to the familiar equation seen in, e.g., Refs.~\cite{Lewin:1995rx,Lin:2019uvt,DelNobile:2021wmp}:
\bea
\label{eq:opticallythinlimit}
\nn \frac{dR}{d\Erec} &=& \bigg(\frac{\rhox}{\mdm}\bigg) \epsilon_{\rm NR} \int_{\rm v_{\rm min}}^{v_{\rm esc}} d^3v v f_{\rm lab}(\Vec{v}) \frac{d\sigmaTx}{d\Erec}~,\\
\frac{d\sigmaTx}{d\Erec} &=& \frac{\mT}{2 \muTx^2 v^2} \sigmaTx S(\Erec)~.
\eea
The differential cross section above corresponds to the one for a DM-nucleus contact interaction that is scalar-scalar (axial-axial) in the non-relativistic limit of spin-independent (spin-dependent) scattering, as canonically assumed in results displayed by experimental searches.

For coherent spin-independent scattering the DM-nucleus cross section $\sigmaTx$ is written in terms of the DM-nucleon cross section $\sigmaNx$ as
\beq
\sigmaTx = \sum_i \beta_i \bigg( \frac{f_p Z_i + f_n (A_i-Z_i)} {f_p}  \bigg)^2 \bigg( \frac{\muTx}{\muNx}\bigg)_i^2  \sigmaNx~, 
\label{eq:SIA4scaling}
\eeq
where $A_i$ and $Z_i$ are the nucleon and atomic number of the target nuclide of isotope $i$ with abundance $\beta_i$, $f_n$ and $f_p$ are the effective DM-neutron and DM-proton couplings, and 
$\muNx$ is the DM-nucleon reduced mass.
We will consider both isospin-conserving ($f_n = f_p$) and isospin-violating ($f_n \neq f_p$) scenarios.
We will also consider a ``no scaling'' scenario as done by recent experiments,
\beq
\sigmaTx = \sigmaNx~,
\label{eq:SInoscaling}
\eeq
corresponding to, {\em e.g.}, strongly-coupled composite DM opaque to the nucleus so that the zero-momentum scattering cross section is geometric~\cite{Digman:2019wdm}.
For spin-dependent scattering,
\beq
\sigmaTx = \sum_i \alpha_i \frac{4}{3} \frac{J_i+1}{J_i} \bigg( \frac{\muTx}{\muNx}  \bigg)_i^2  \sigmaNx~, 
\label{eq:SDXS}
\eeq
where $\alpha_i$ are the number fractions of non-zero spin isotopes of spin $J_i$. 

\subsection{Scaling trends}
\label{subsec:scalingtrends}

Some aspects of the single-scatter ceiling are already apparent from the treatment above.
For one, Eq.~\eqref{eq:NevSSmaster} implies that for a given DM mass there are two values of the scattering cross section that produce the same number of single-scatter events.
One of them is from the familiar optically thin limit, often quoted by experiments as the exclusion cross section. 
In this case, for $\mdm \gg \mT$ where $\muTx \to \mT$, $\sigmaTxeff$ in Eq.~\eqref{eq:sigmaTxeff} is $\mdm$-independent and we get a scaling relation for it linear with $\mdm$:
\beq
_b\sigmaTxeff = \bigg(\frac{1}{\nT L_{\rm ave}}\bigg) \bigg(\frac{N_{\rm ev}^{\rm SS}}{\epsilon_{\rm NR}}\bigg) \bigg(\frac{\mdm}{\bar m}\bigg)~,
\label{eq:sigTxeffvmdm-bot}
\eeq
where $\bar m = \pi \rho_\chi \Rdet^2 \bar v \texp$.
The second solution is obtained near the optically thick limit by setting $k = 1$ in Eq.~\eqref{eq:probmultiplic}, giving a logarithmic scaling with $\mdm$:
\beq
_t\sigmaTxeff \simeq \bigg(\frac{1}{\nT L_{\rm ave}}\bigg) \bigg[
 \log \bigg(\frac{\bar m}{\mdm}\bigg)
- \log \bigg(\frac{N_{\rm ev}^{\rm SS}}{\epsilon_{\rm NR}}\bigg) 
\bigg]~,
\label{eq:sigTxeffvmdm-top}
\eeq
where we have neglected a log($\Nhit$) term.

Experimental exclusion cross sections correspond to some fixed value of $N_{\rm ev}^{\rm SS}$, thus the usually quoted limits on $\sigmaNx$ (as derived from Eq.~\eqref{eq:sigTxeffvmdm-bot}) increase linearly with $\mdm$.
Physically, increasing $\mdm$ reduces the DM flux, so the $\sigmaNx$ sensitivity reduces proportionally.
On the other hand, the ceiling on $\sigmaNx$ decreases logarithmically with $\mdm$.
This is again because the DM flux reduces with increasing $\mdm$, but in a subtler way.
At low $\mdm$ the integrated flux $\Phi$ is large, and single-scatter events via Poisson fluctuations are obtainable deeper into the optically thick regime.
As $\mdm$ is increased, the smaller $\Phi$ implies it is more probable to obtain single scatters via Poisson fluctuations at lower cross sections. 

It is also interesting to note the scalings of Eqs.~\eqref{eq:sigTxeffvmdm-bot} and \eqref{eq:sigTxeffvmdm-top} with the detector mass and live time.
As $\Mdet = \pi \rhoT \Rdet^2 L_{\rm ave}$, we have 
\bea
 _b\sigmaTxeff &\propto& (\Mdet \texp)^{-1}~,\\
\nn  _t\sigmaTxeff &\propto& \Mdet^{-1/3} \log (\Mdet^{2/3} \texp)~.
\label{eq:sigTxeffvMdettexp}
\eea
As expected, the sensitivity of $_b\sigmaTxeff$ improves proportional to the exposure $\Mdet \texp$.
But the sensitivity of $_t\sigmaTxeff$ scales more non-trivially.
As $\Mdet$ is increased, $_t\sigmaTxeff$ {\em decreases} as $\Mdet^{-1/3}$ falls faster than the rise of $\log \Mdet^{2/3}$.
Yet as $\texp$ is increased, $_t\sigmaTxeff$ {\em increases} logarithmically.
Thus the notion of ``exposure'' $\Mdet \texp$ is not particularly useful in determining single-scatter ceilings (and the neutrino roof), but rather one must specify $\Mdet$ and $\texp$ separately.
Therefore two different detector configurations ($\Mdet$, $\texp$) of the same target material and with the same exposure $\Mdet \texp$, such as achievable in DARWIN/XLZD and PANDAX-xT, become important for reaching complementary regions of single-scatter ceilings.
Physically, we obtain this behaviour for $_t\sigmaTxeff$ because the larger the detector is, the lower is the cross section that forms the single-scatter/multiscatter boundary.
Also, the longer the run-time is, the higher is the integrated DM flux, hence  (logarithmically) higher cross sections may be reached by a single-scatter search via Poisson fluctuations in multiplicity.

These scalings also tell us that the single-scatter ceilings cannot be pushed appreciably higher by increasing the exposure, unlike the single-scatter floors that have been pushed down by many orders of magnitude over four decades by steady increase of exposure.
Finally, since the $t_{\rm exp}$ of experiments are generally comparable to each other (at $\Oc$(yr)) whereas $\Mdet$ has been increasing by orders of magnitude, the single-scatter ceilings of experiments have generally moved down in cross sections.
This means that older experiments already rule out single-scatter ceiling cross sections of newer experiments -- but only up to a certain DM mass.
As exposures increase, DM integrated fluxes $\propto \mdm^{-1}$ too increase, and therefore higher $\mdm$ may be reached.
For this reason, we will display results that focus on ultra-heavy DM to highlight the new parameter space constrained below single-scatter ceilings.
We turn to this task in the next section.

\section{Results}
\label{sec:results}

\subsection{Statistics}

Assuming that event counts follow Poisson statistics and that background events follow a Gaussian
distribution with mean $N_{\rm B}$ and standard deviation $\sigma_B$, the 90\% C.L. limit $N_{\rm exp}^{\rm 90CL}$ for $N_{\rm obs}$ events observed in the selected range of $\Erec$ is given by
\begin{widetext}
\bea
\label{eq:def:90CL}
&& \int_0^\infty dB \frac{\Gamma(N_{\rm obs} + 1, N_{\rm exp}^{\rm 90CL} + B)}{N_{\rm obs}!} \zeta_B \exp \bigg[-\frac{(B-N_{\rm B})^2}{2\sigma_B^2} \bigg] = 0.10~, \\
\nn && \zeta_B^{-1} = \sqrt{\frac{\pi}{2}} \sigma_B \bigg(1 + {\rm erf}\bigg( \frac{N_{\rm B}}{\sqrt{2} \sigma_B} \bigg) \bigg)~, 
\eea
\end{widetext}
where $\Gamma$ is the incomplete gamma function and $\zeta_B$ normalizes the Gaussian.
As we will see in Sec.~\ref{subsec:nuroofplots}, backgrounds from atmospheric neutrinos dominate over other sources for future detectors.
Assuming a standard 20\% uncertainty on the flux prediction~\cite{nufloor:atmosnuuncert:Honda2011}, the total uncertainty on the background is obtained as $\sigma_B = \sqrt{N_{\rm B} + (0.2 N_{\rm B})^2}$ for $\NBG \gg 1$.
In the limit of $\sigma_B \to 0$, as is the case with $\NBG \lsim \Oc(1)$, the Gaussian in Eq.~\eqref{eq:def:90CL} becomes a delta function and the equation becomes 
\beq
\frac{\Gamma(N_{\rm obs} + 1, N_{\rm exp}^{\rm 90CL} + N_{\rm B})}{N_{\rm obs}!} = 0.10~, 
\label{eq:def90CL:noBGerror}
\eeq
which would then give, for $N_{\rm B} = 0$ and  $N_{\rm obs}$ = \{1, 2, 3, 4, ...\},  familiar values of $N_{\rm exp}^{\rm 90CL}$ = \{2.3, 3.9, 6.7, 8,...\}.
Due to the low statistics in our problem, setting limits using Eqs.~\eqref{eq:def:90CL} and \eqref{eq:def90CL:noBGerror} gives results in excellent agreement with those reported by direct searches.
We will use Eq.~\eqref{eq:def:90CL} to also identify DM sensitivity regions for future detectors with large exposures and appreciable backgrounds, which we again find gives results that are very close to ones obtained with binned profile likelihood methods in the literature~\cite{nufloor:OHare:2020lva,nufloor:GaspertGiampaMorrissey:2021gyj}.
We note here that the PANDAX-xT collaboration~\cite{PandaX-xT:2024oxq} derives future sensitivities with an even simpler method than ours, namely, a Gaussian statistics-based cut-and-count.

\begin{figure*}[ht]
    \centering
    \includegraphics[width=.47\textwidth]{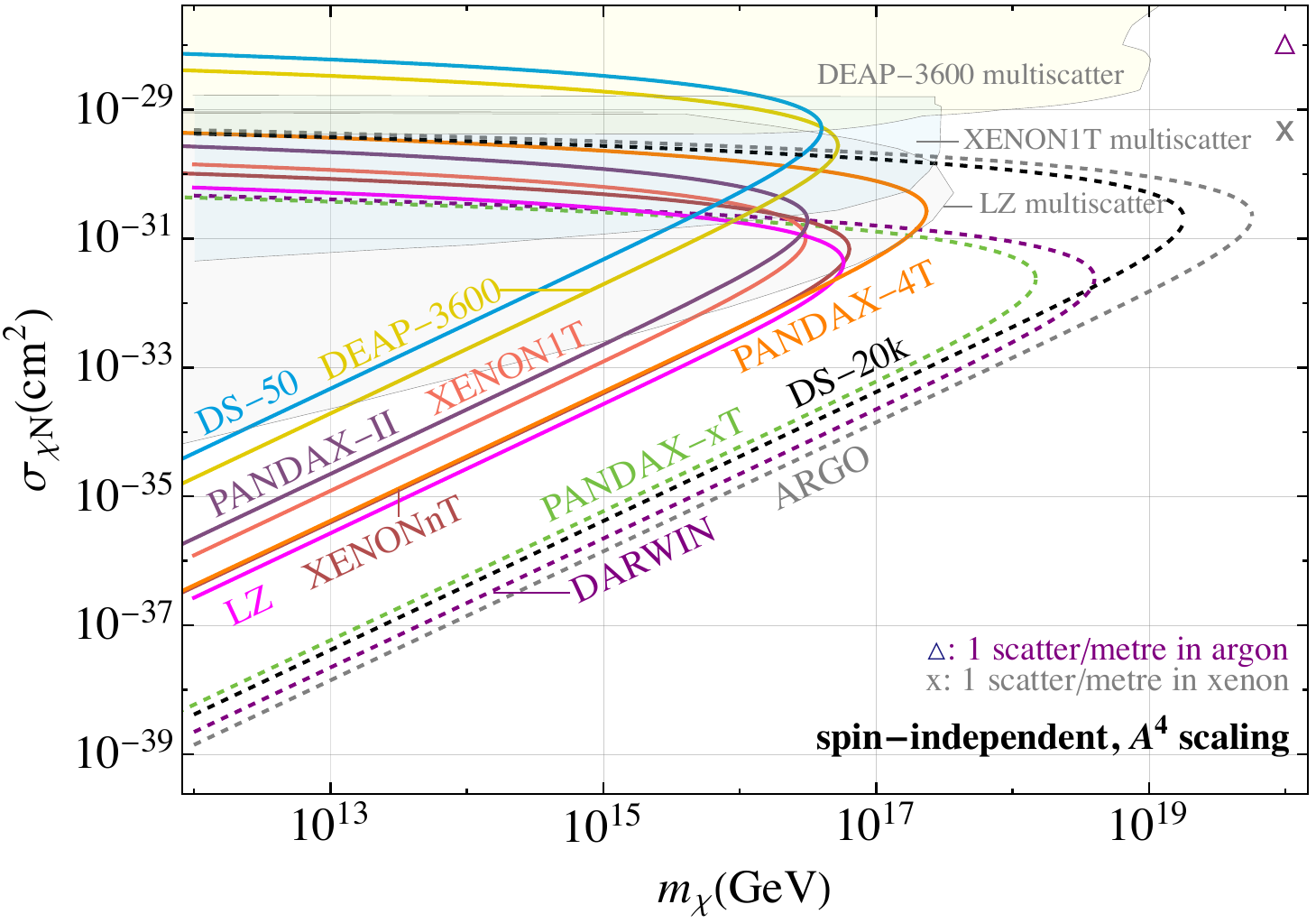}  \includegraphics[width=.47\textwidth]{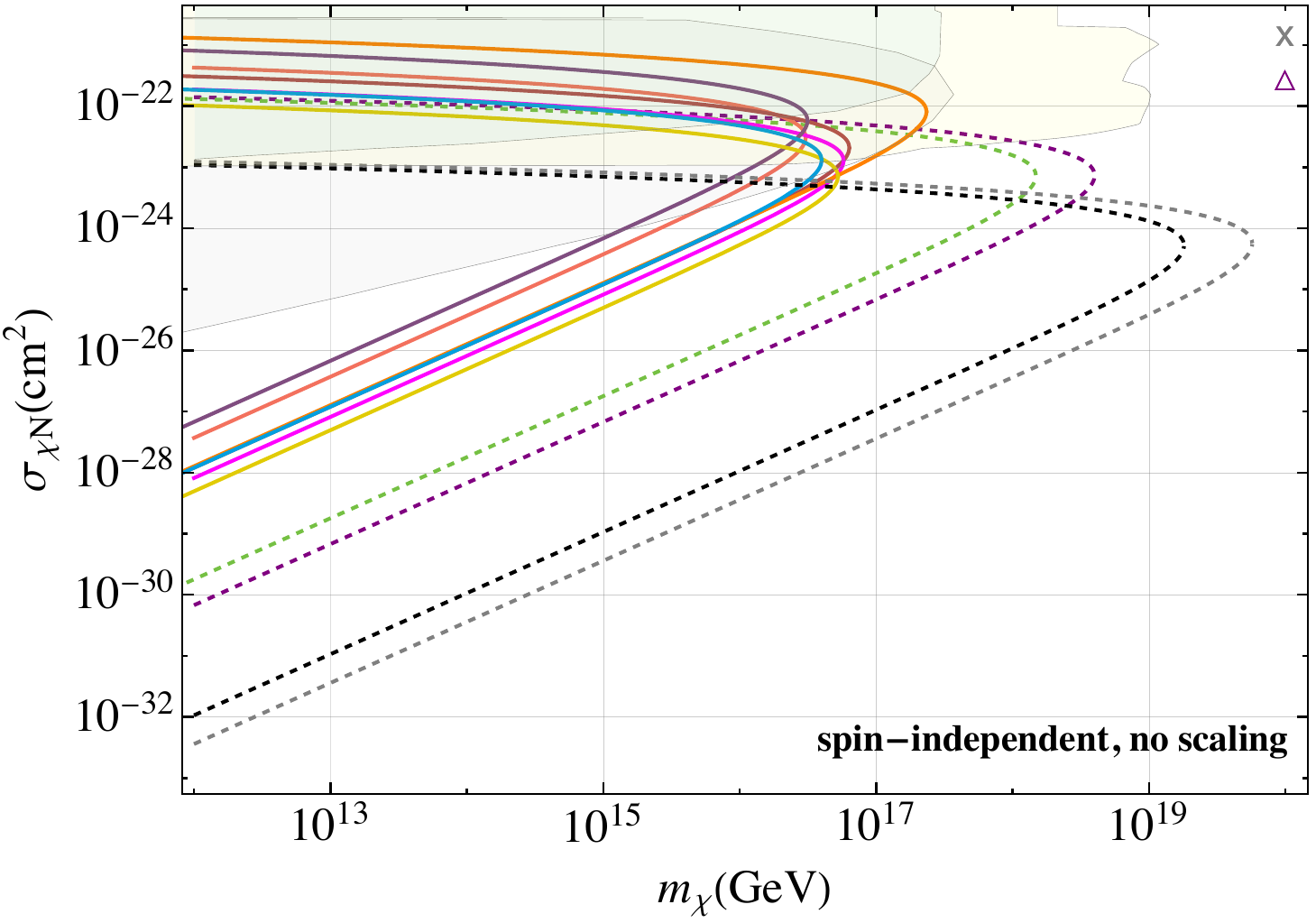} \\
    \includegraphics[width=.47\textwidth]{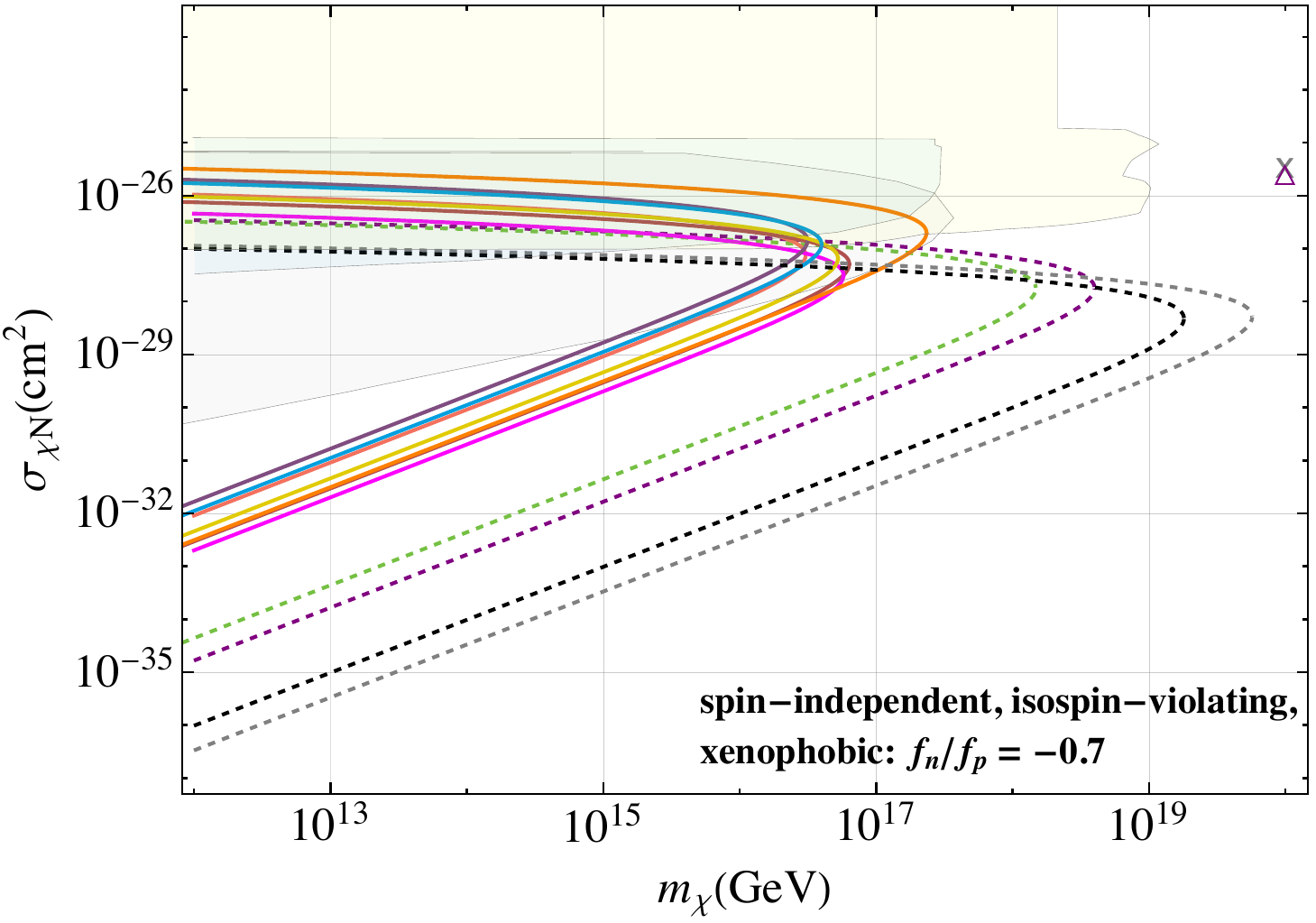}  \includegraphics[width=.47\textwidth]{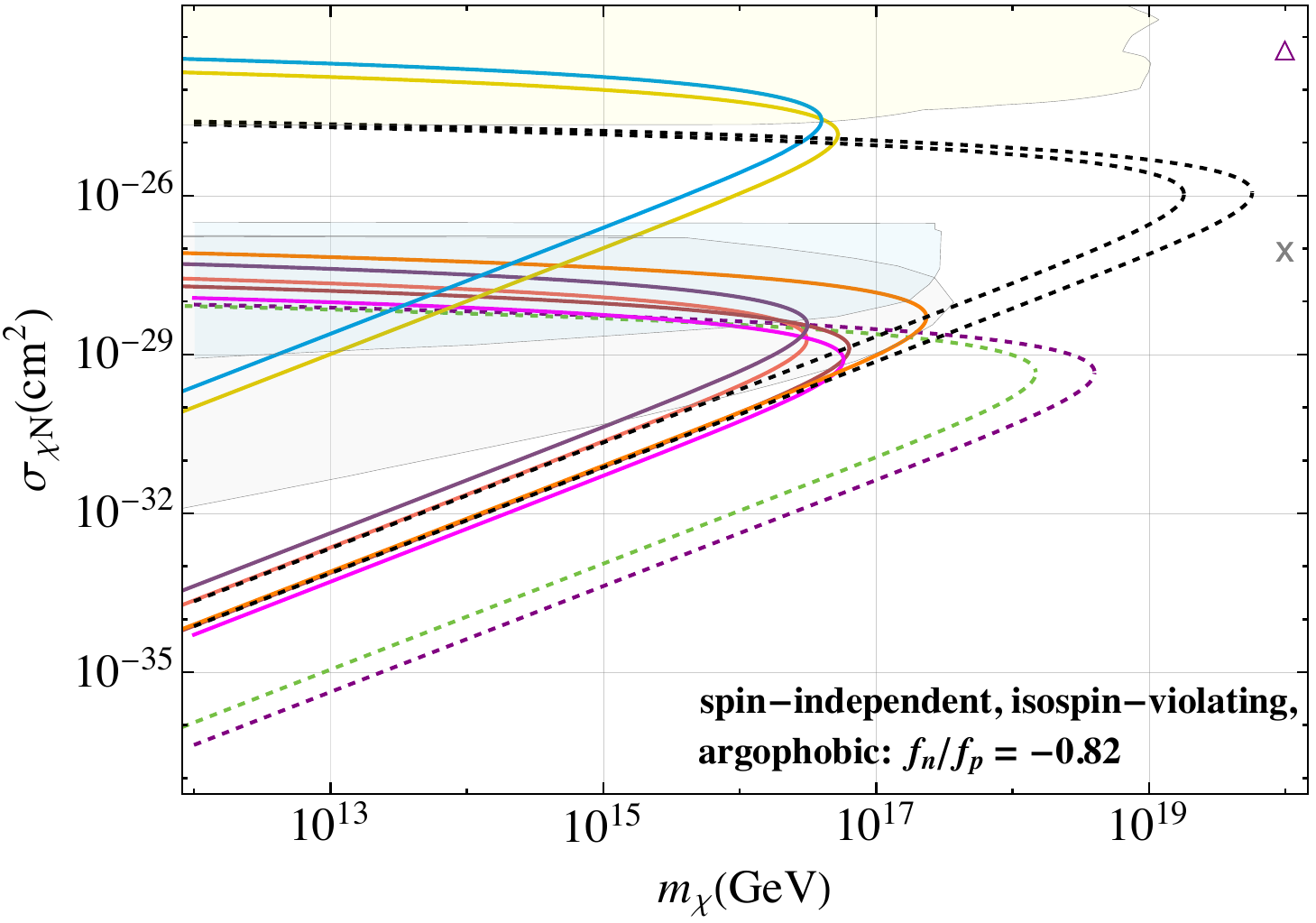} \\
    \includegraphics[width=.47\textwidth]{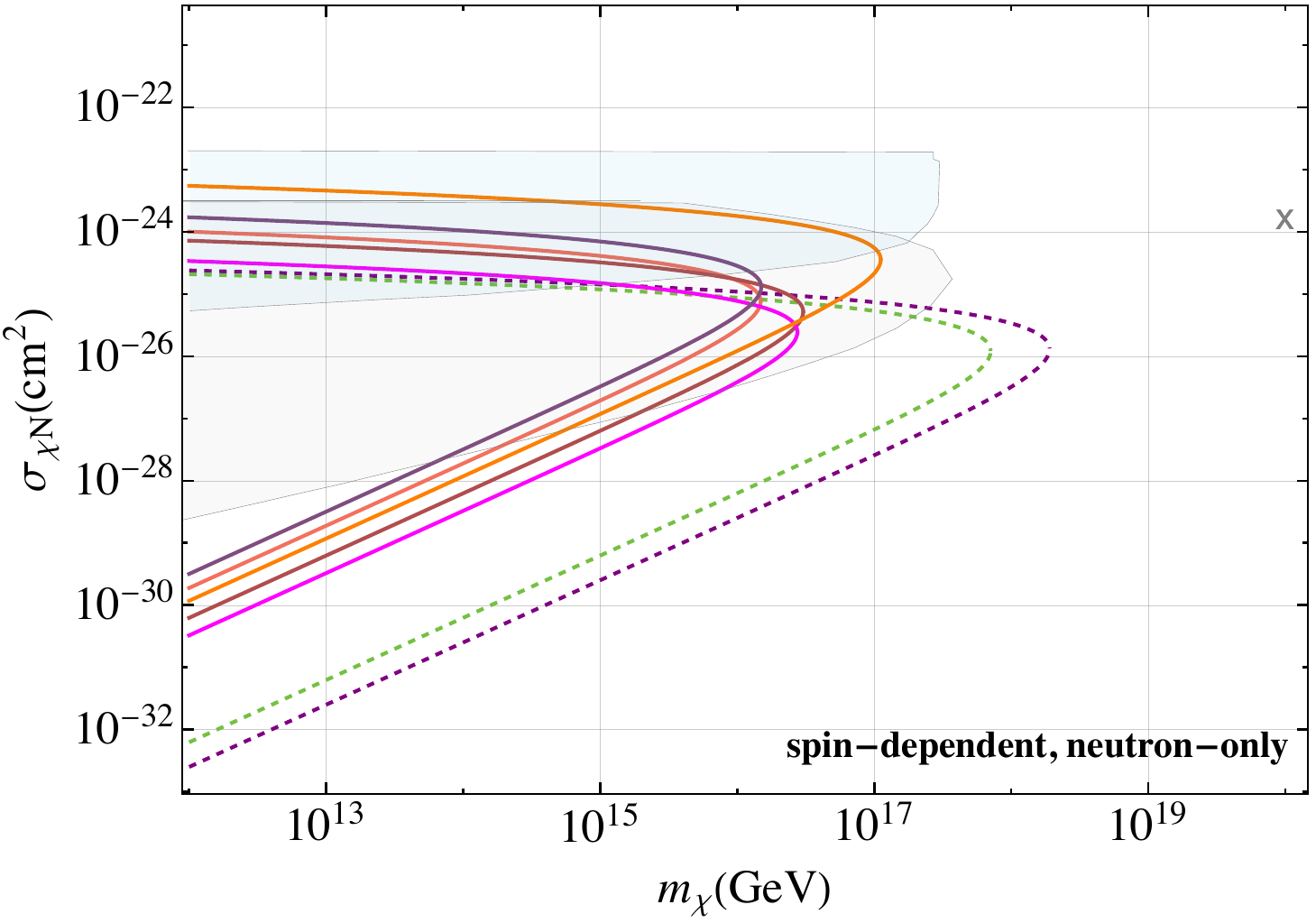}  \includegraphics[width=.47\textwidth]{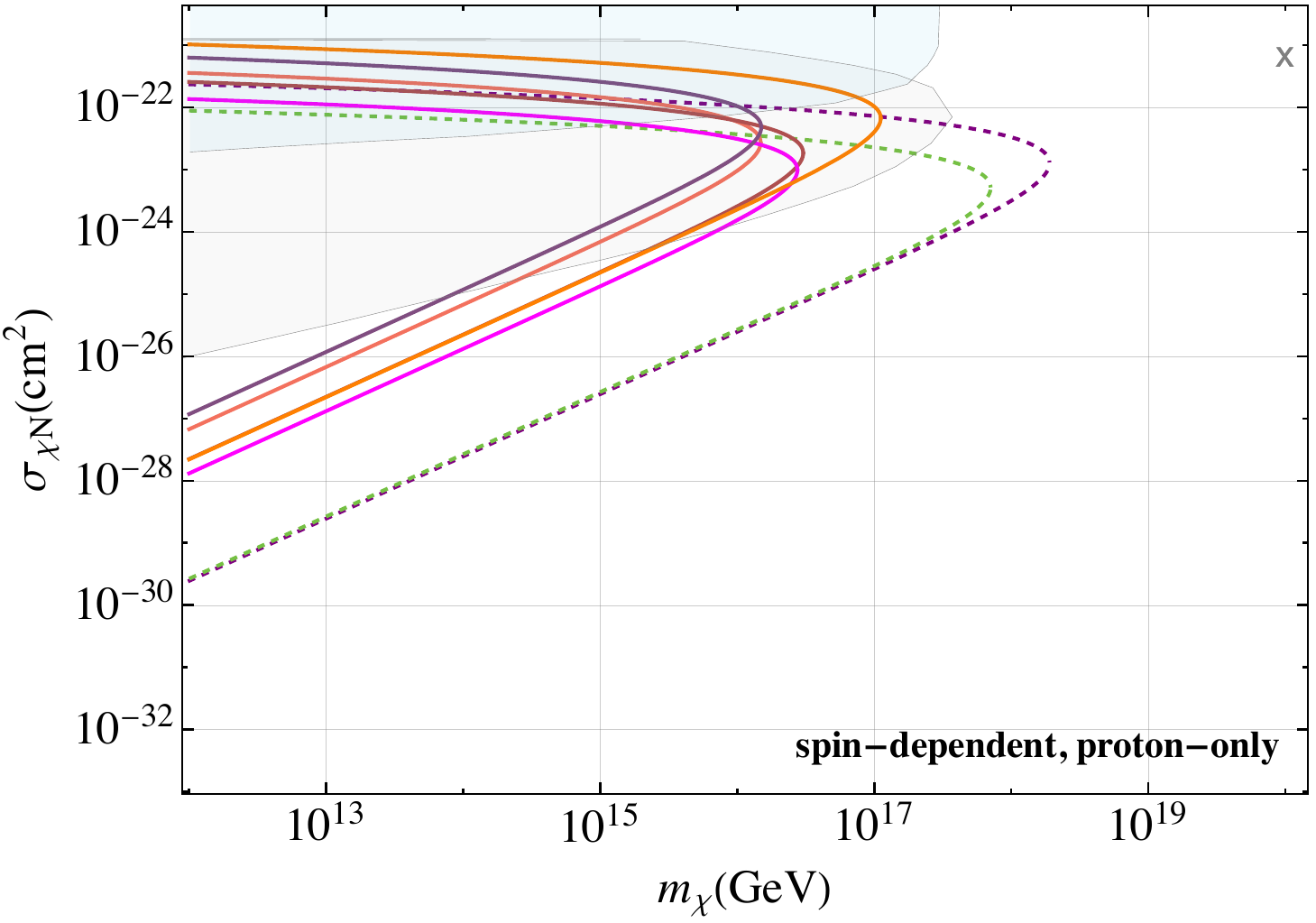}
    \caption{90\% C.L. single-scatter limits (solid curves) and projections (dashed curves) on DM-nucleon cross sections and DM masses for the various scattering scenarios described in Sec.~\ref{subsec:masterformulae}, obtained using Table~\ref{tab:detectordetails} and Eqs.~\eqref{eq:def:90CL} \& \eqref{eq:def90CL:noBGerror}.
    The highlight here is single-scatter ceilings that are typically not shown in direct detection exclusion regions. 
    Also displayed are current limits from multiscatter searches in shaded regions, and cross sections for which one scatter per metre of detector material are expected. 
    See Sec.~\ref{subsec:limits} for further details.}
    \label{fig:limits}
\end{figure*}

\subsection{Current limits and next-generation projections}
\label{subsec:limits}

In Fig.~\ref{fig:limits} we display our results in the space of DM-nucleon scattering cross section and DM mass by setting $N_{\rm exp}^{\rm 90CL}$ in Eqs.~\eqref{eq:def:90CL} and \eqref{eq:def90CL:noBGerror} to $N_{\rm ev}^{\rm SS}$ in Eq.~\eqref{eq:NevSSmaster} and using the information in Table~\ref{tab:detectordetails}.
We take the detection efficiency $\epsilon_{\rm NR}$ as constant in the range of $[E_{\rm R, min}, E_{\rm R, max}]$ specified.
This is justified since, as seen in the experimental references in Table~\ref{tab:detectordetails}, this efficiency typically rises sharply with $\Erec$ below $E_{\rm R, min}$, stays nearly constant for $E_{\rm R, min} \lsim \Erec \lsim E_{\rm R, max}$, and declines sharply with $\Erec$ above $E_{\rm R, max}$.
For the future experiments DARWIN/XLZD, PANDAX-xT, DarkSide-20k and Argo, $\epsilon_{\rm NR}$ must be interpreted as the nuclear recoil (NR) acceptance, which comes at the cost of a desired level of electron recoil (ER) rejection. 
This point will be taken up in more detail in the next sub-section.

The top four panels show spin-independent limits and projections for four scenarios: 
isospin-conserving with $f_n = f_p$ in Eq.~\eqref{eq:SIA4scaling} so that $\sigmaTx \propto A^4 \sigmaNx$, 
isospin-violating \& xenophobic with $f_n/f_p = -0.7~(\simeq -Z_{\rm Xe}/(A_{\rm Xe} - Z_{\rm Xe}))$, 
isospin-violating \& argophobic with $f_n/f_p = -0.82~(\simeq -Z_{\rm Ar}/(A_{\rm Ar} - Z_{\rm Ar}))$,
and the case of no $A$-scaling in Eq.~\eqref{eq:SInoscaling}.
The bottom two panels show spin-dependent limits with neutron-only and proton-only limits (Eq.~\eqref{eq:SDXS}).
These panels only show xenon detector limits as argon has no spin-carrying isotopes.

In all panels we see both the upper bound (floor) and lower bound (ceiling) of single-scatter searches, corresponding to the two solutions of Eq.~\eqref{eq:NevSSmaster} in Eqs.~\eqref{eq:sigTxeffvmdm-bot} and \eqref{eq:sigTxeffvmdm-top}.
The turnaround point occurs where these two solutions are equal, which is at
\beq
 m_{\rm \chi}^{\rm max} \simeq \pi \rho_\chi \Rdet^2 \bar v \texp \epsilon_{\rm NR}/N_{\rm exp}^{\rm 90CL}~.
 \label{eq:mxturnaround}
\eeq
In the top two panels the single-scatter ceilings for a given target material are lower for higher $\Mdet$, as described in Sec.~\ref{subsec:scalingtrends}.
In the top right panel, the xenon single-scatter ceilings are generally higher than the argon ceilings for comparable detector masses.
This is because the event rates in xenon are generally more suppressed by the Helm form factor in Eq.~\eqref{eq:sigmaTxeff} in the $\Erec$ range of interest: xenon being a larger nucleus than argon, coherent scattering is harder to achieve.
In the top left panel, however, the xenon single-scatter ceilings are lower than the argon ceilings due to the $A^4$ scaling.
Thus while xenon is a better element than argon for probing small cross sections in this case thanks to its larger $A^4$-enhancement, argon is better for reaching large cross sections.
The $A^4$-scaling itself may break down for $\sigmaNx \gsim 10^{-32}-10^{-30}$~cm$^2$ due to failure of the Born approximation~\cite{Digman:2019wdm}, but may still hold for dark nuclei~\cite{DEAP:MS:2021raj}.
In any case we show results with $A^4$-scaling for all values of $\sigmaNx$ for comparison with and across experiments. 
Our results for XENON1T and LZ are in close agreement with the high-mass single-scatter floors and ceilings reported by the collaborations themselves~\cite{XENON1T:MSSSprojexn:Clark:2020mna,XENON1T:MS:2023iku,LZ:MS:2024psa}, who had obtained multiplicity distributions with Monte Carlo simulations to capture their respective detector geometries.
This agreement justifies our approach of using Poisson distributions in multiplicities with spherical detectors.

\begin{figure*}[ht]
    \centering
    \includegraphics[width=.47\textwidth]{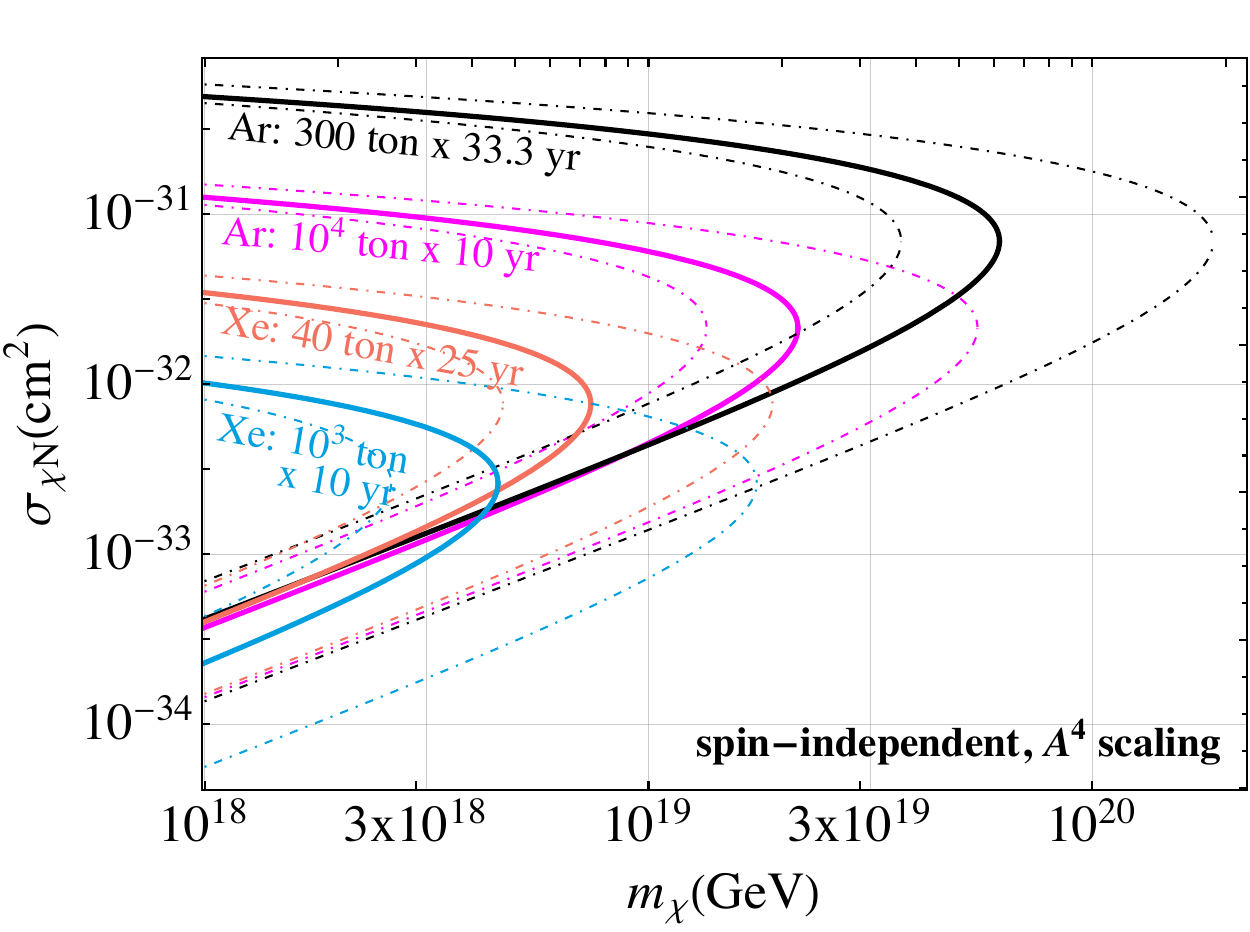}  \includegraphics[width=.47\textwidth]{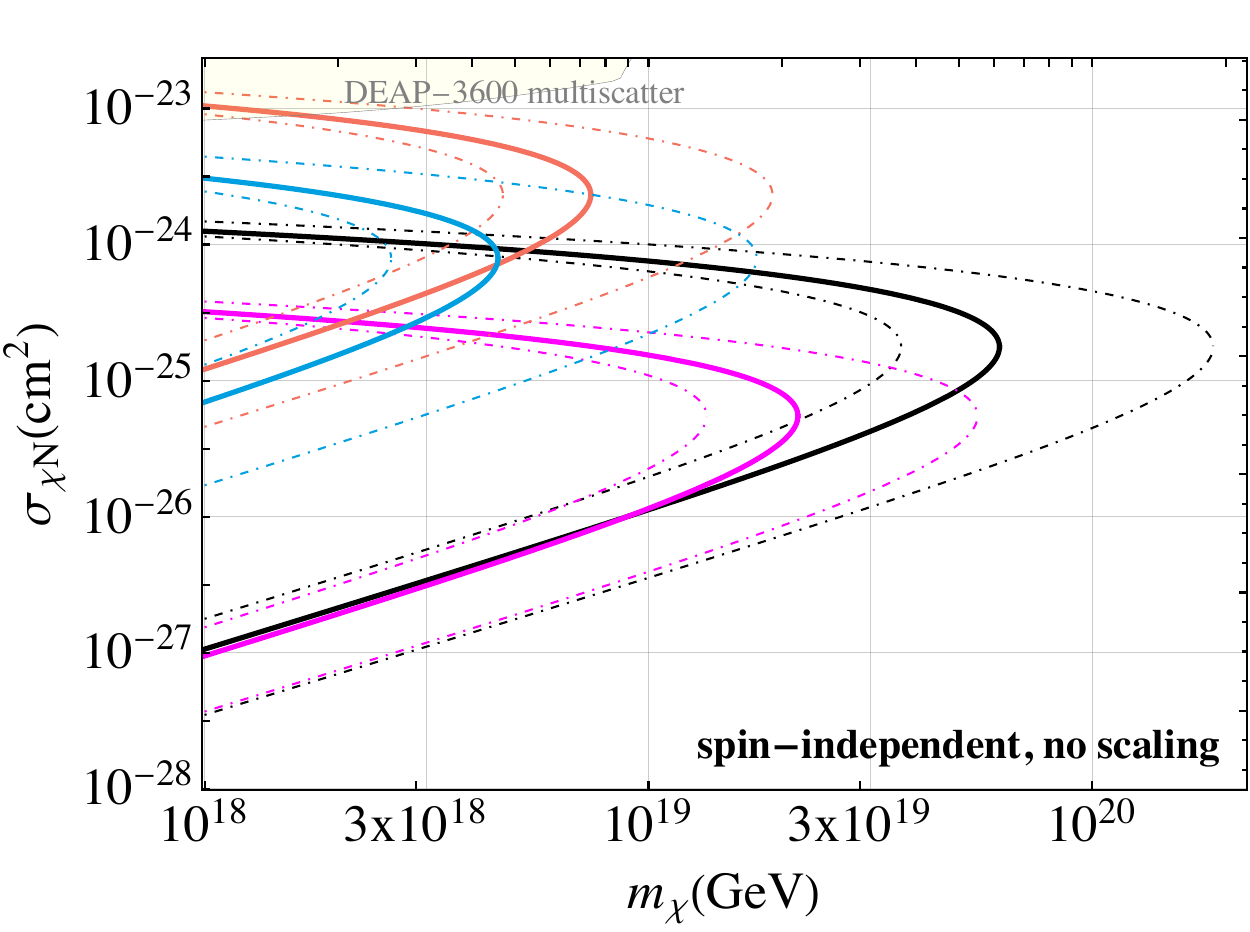} \\
    \includegraphics[width=.47\textwidth]{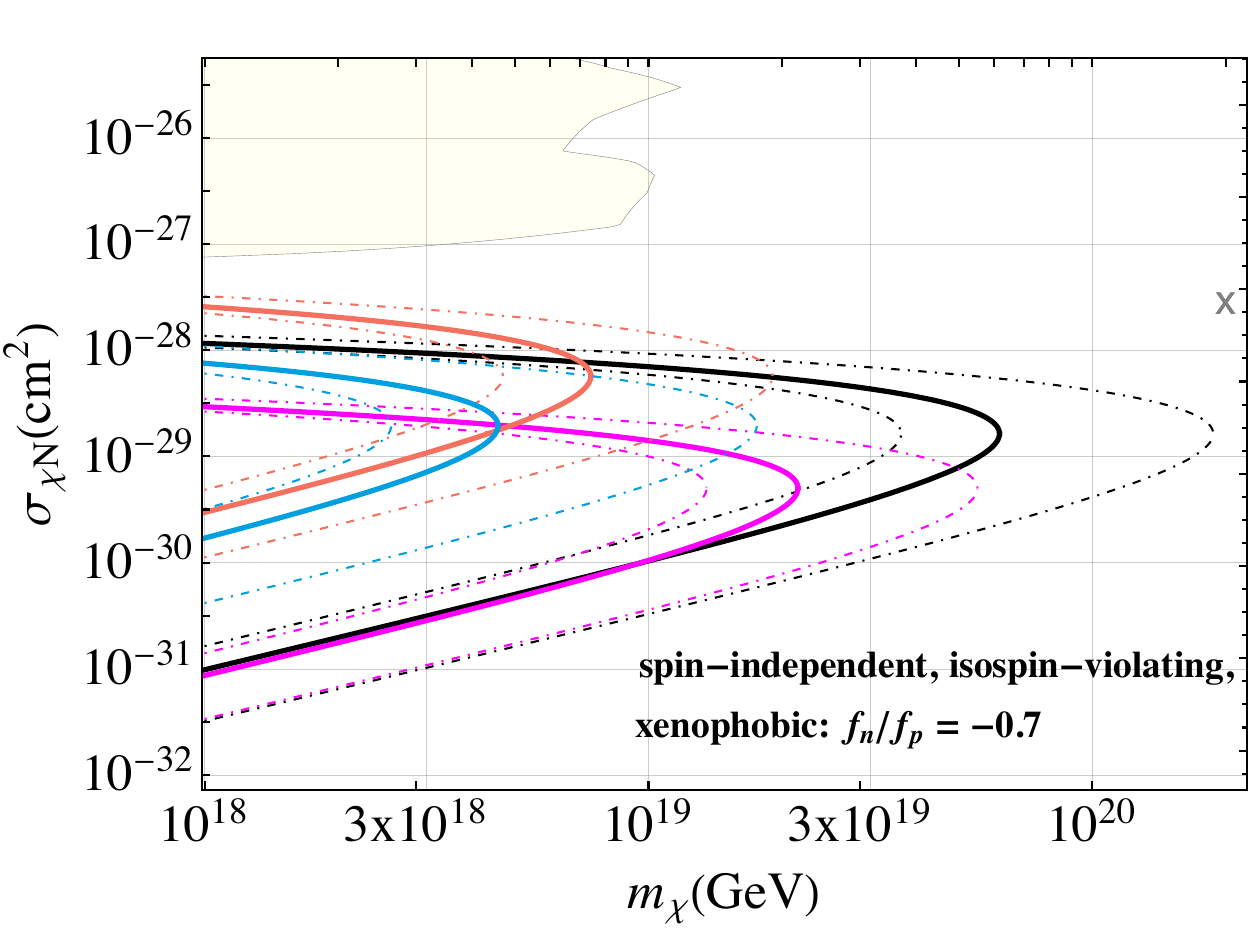}  \includegraphics[width=.47\textwidth]{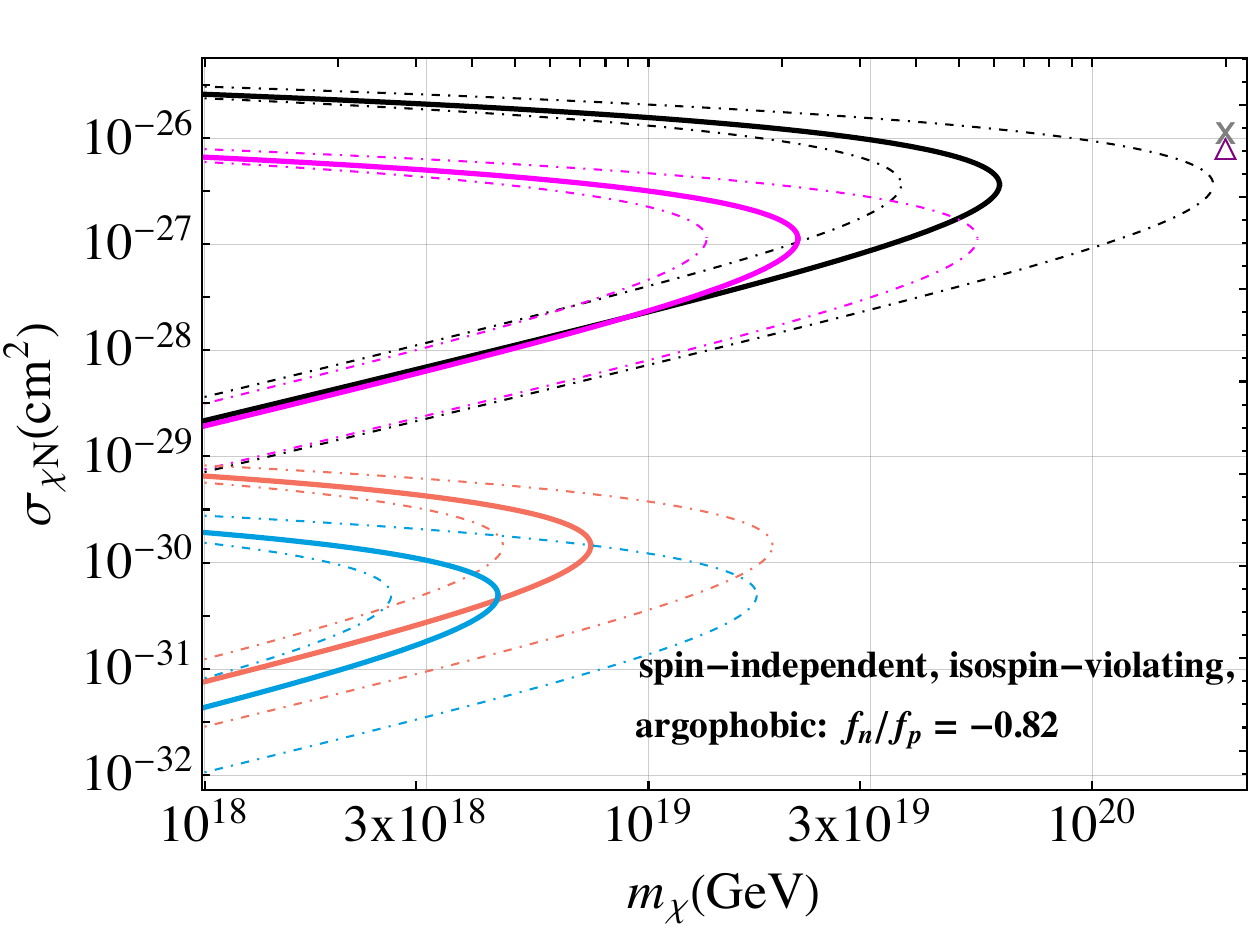} \\
    \includegraphics[width=.47\textwidth]{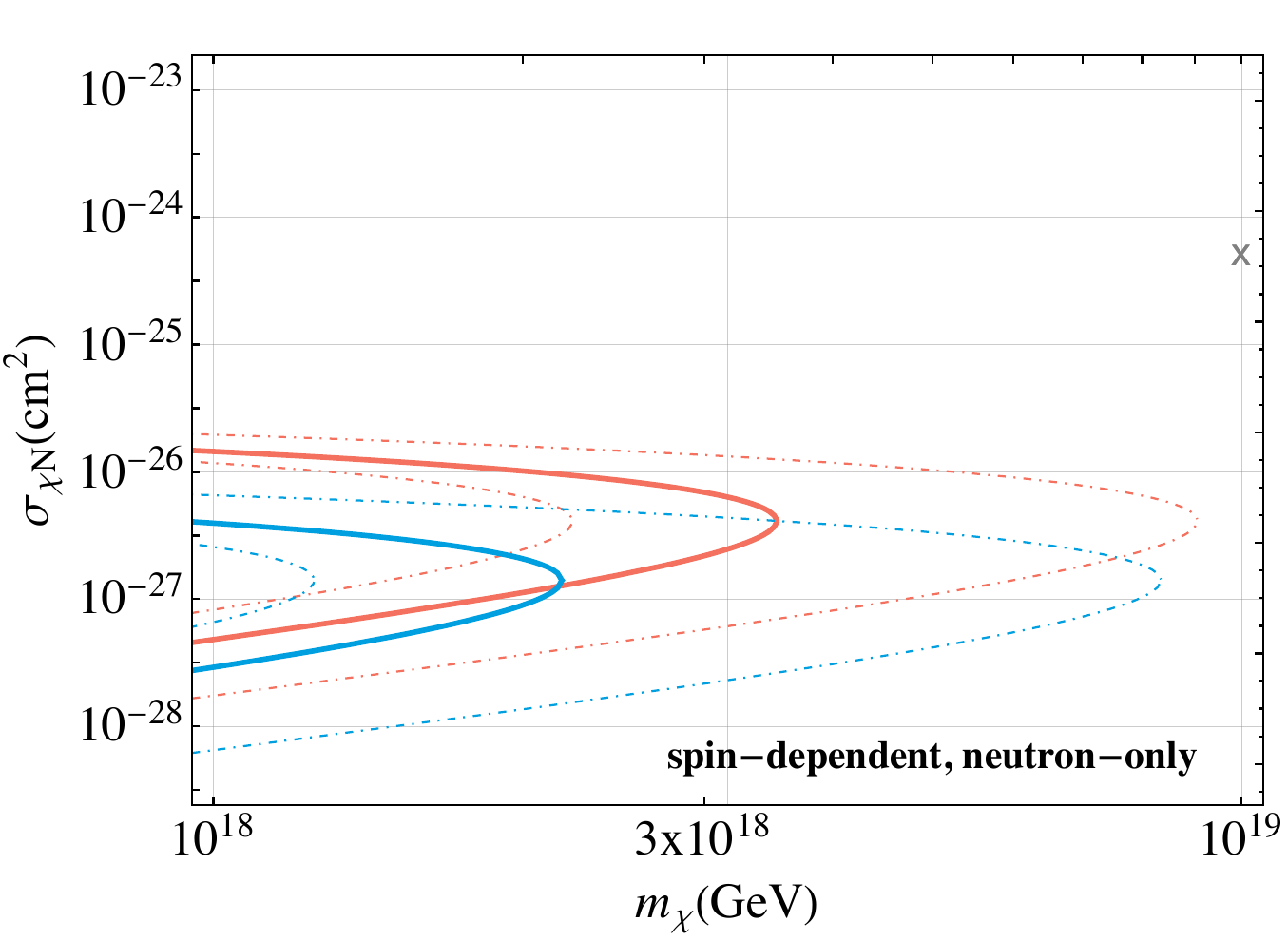}  \includegraphics[width=.47\textwidth]{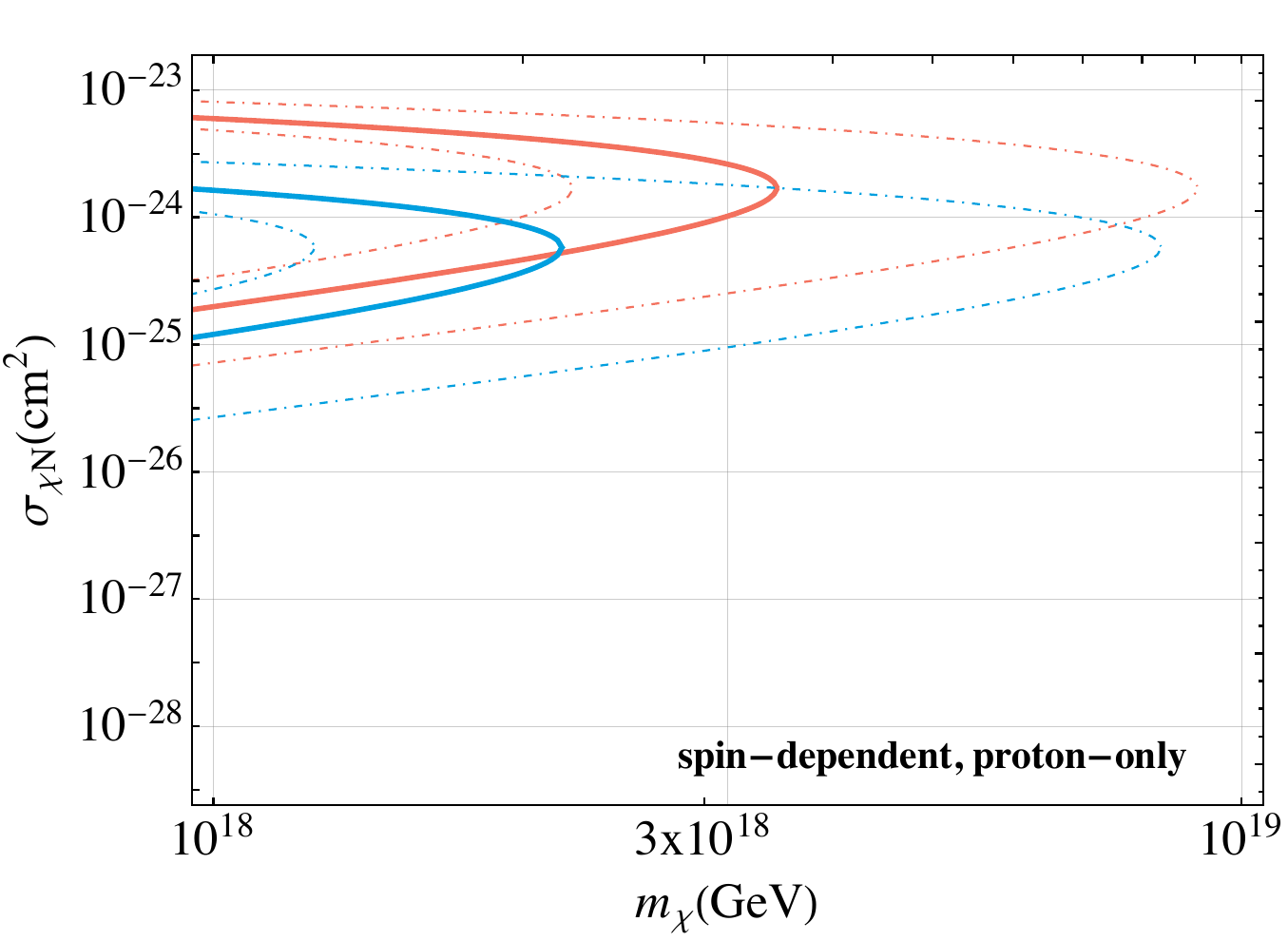}
    \caption{90\% C.L. single-scatter reaches in the presence of large atmospheric neutrino backgrounds at hypothetical large-exposure detectors listed in Table~\ref{tab:detectordetails}.
    The dot-dashed curves account for fluctuations in the number of events expected to be observed.
    The detector fiducial mass and live time are indicated; variations in both while keeping the exposure $\Mdet \times \texp$ unchanged would alter the single-scatter ceiling: see Appendix~\ref{app:Mdetvariedroofs}.
    As in Fig.~\ref{fig:limits}, multiscatter limits and cross sections giving one scatter per metre are shown.
    See Sec.~\ref{subsec:nuroofplots} for further details.}
    \label{fig:nuroof}
\end{figure*}

In all panels we show in shaded regions limits from multiscatter searches at DEAP-3600~\cite{DEAP:MS:2021raj}, XENON1T~\cite{XENON1T:MS:2023iku}, and LZ~\cite{LZ:MS:2024psa}.
For reference, we also mark (with ``$\Delta$'' and ``x'') the cross section at which we expect one scatter per metre in xenon and argon, assuming the $\Erec$ range of DARWIN/XLZD and Argo respectively as given in Table~\ref{tab:detectordetails}.
These points give an idea of where transitions from single-scatter to multiscatter regimes may be expected.
They also roughly follow the relative positions of xenon vs argon single-scatter ceilings.
We see that single-scatter exclusion regions near their ceilings generally overlap with multiscatter exclusion regions, as expected from Poisson fluctuations in multiplicity.
This underlines the significance of single-scatter ceilings at future detectors, which would reach DM masses higher than excluded by current multiscatter searches.
The overlap of exclusion regions is also important cross-confirmation from two classes of searches using the same detector material; see Appendix~\ref{app:searchmethods} for how these searches were performed.
We do not display multiscatter null results from DAMA~\cite{DAMA1999}, Edelweiss~\cite{EdelweissCDMSAlbuquerqueBaudis2003}, CDMS~\cite{EdelweissCDMSAlbuquerqueBaudis2003,KavanaghSuperheavy:2017cru},
PICO-60~\cite{PICO:MS:Broerman2022},
MAJORANA DEMONSTRATOR~\cite{XENON1T:MSSSprojexn:Clark:2020mna},
the Chicago detector of Ref.~\cite{CollarBeacomCappiello2021},
and searches in mica~\cite{mica:pricesalamon1986} and
plastic etch detectors~\cite{PlasticEtch:Bhoonah2020fys}, 
either because their exclusion regions are subsumed within the limits we show or because they constrain cross sections larger than shown here.
Similarly the cross section ceiling from the Earth overburden is too high to show here~\cite{KavanaghSuperheavy:2017cru,Bramante:2018qbc,Bramante:2018tos}.
Compendia of these limits may be found in Refs.~\cite{KavanaghSuperheavy:2017cru,gascloudcompendium:Bhoonah:2018gjb,PlasticEtch:Bhoonah2020fys,snowmass:Carney:2022gse}.

The xenon and argon single-scatter ceilings are closely spaced for xenophobic DM in the middle left panel.
This is a result of the xenon $\sigmaNx$ sensitivities being relatively pushed up compared to the top left panel, where they enjoyed $A^4$ scaling.
The xenon and argon limits -- both floor and ceiling -- are dramatically separated for argophobic DM in the middle right panel.
This is because for a given $\sigmaNx$, the relative scattering rate on argon vs xenon drops more sharply for this case than it does in the xenophobic case.
This in turn is because xenon is composed of multiple isotopes of comparable abundances, so that fixing $f_n/f_p$ to some value does not suppress $\sigmaTx$ too strongly in the summation of Eq.~\eqref{eq:SIA4scaling}, whereas argon is predominantly made of $^{40}$Ar and $\sigmaTx$ is therefore greatly suppressed for $f_n/f_p = -0.82$.
As pointed out in Refs.~\cite{Xephobic:Feng:2013vod,Xephobic:Yaguna:2019llp} the case of xenophobic DM highlights the importance of the less-heeded liquid argon program of direct detection, as it is a scenario where argon detectors may set better upper bounds (floors) on $\sigmaNx$.
Now we see that the case of argophobic DM lends further weight to the liquid argon program, as argon detectors probe cross sections much higher than reachable by xenon detectors.
We believe this is an important science case for the commissioning of the future Argo detector.

Yet another reason for why detectors made of different elements are important in their own right is that the scalings from per-nucleon to per-nucleus cross sections are determined by the unknown scattering potential, and could be virtually anything for both elementary and composite DM~\cite{Digman:2019wdm}.
This means comparisons of the per-nucleon cross section across experiments using different nuclear targets are generally unreliable.
In this light, the xenon and argon per-nucleus limits shown in the top right panel of Fig.~\ref{fig:limits} should be read independent of each other. 

Moving on to spin-dependent limits in the bottom two panels, the relative ordering of the xenon-scattering limits in each panel unsurprisingly follows that of the spin-independent cases.
That the proton-only limits are weaker than the neutron-only limits by two orders of magnitude is a reflection of the corresponding structure factor being smaller by $\Oc(10^2)$ across recoil energies~\cite{spindepstrucfac:Klos:2013rwa}. 
One may also get a feel for this by looking at the zero-momentum spin expectation values $\{\langle S_n \rangle, \langle S_p \rangle \}$ in Ref.~\cite{spindepstrucfac:Klos:2013rwa}, to wit, $\{0.33, 0.01\}$ for $^{129}$Xe and
$\{-0.27, -0.01\}$ for $^{131}$Xe.
We note here that spin-dependent limits were not displayed by LZ in the work where spin-independent multiscatter and single-scatter limits in the high mass region were derived~\cite{LZ:MS:2024psa}.
The LZ limits in the bottom panels of Fig.~\ref{fig:limits} are our recasting of the WIMP search in Ref.~\cite{LZ:SS:2022lsv}.

\subsection{Projections to very large exposures}
\label{subsec:nuroofplots}

In Fig.~\ref{fig:nuroof} we show 90\% C.L. sensitivities for the four spin-independent and two spin-dependent scenarios seen in Fig.~\ref{fig:limits}, for hypothetical exposures larger than those that will be achieved in DARWIN/XLZD, PANDAX-XT, DarkSide-20k, and Argo.
Once again we show for reference current multiscatter limits (of which only DEAP-3600 results are visible in the $\mdm$ range shown) and cross sections for which unit per-metre multiplicity are expected.
For xenon detectors, we consider $10^3$~ton-yr (40 ton $\times$ 25 yr) and $10^4$~ton-yr (10$^3$ ton $\times$ 10 yr), and for argon, $10^4$~ton-yr (300 ton $\times$ 33.3 yr) and $10^5$~ton-yr (10$^4$ ton $\times$ 10 yr).
We specify the fiducial mass and live time separately as they play different roles in setting the single-scatter ceilings as discussed below Eq.~\eqref{eq:sigTxeffvmdm-top}; in Appendix~\ref{app:Mdetvariedroofs} we show some single-scatter ceilings for other ($\Mdet, \texp$) configurations keeping the exposure fixed to the above values.
The detector configurations mentioned here are realistic extensions of what is currently believed to be achievable with DARWIN/XLZD (40 ton $\times$ 5 yr), PANDAX-xT (34.2 ton $\times$ 5.85 yr), and Argo (300 ton $\times$ 10 yr).
Procuring $\Oc(10^3)$ tonnes of xenon, a goal shared by $0\nu\beta\beta$ experimenters, may be done in stages driven by the demands of semiconductor manufacturing, and medical, lighting and spacecraft propulsion research, as well as via developing techniques such as Direct Air Capture~\cite{Xekton:Avasthi:2021lgy,*Xekton:Anker:2024xfz}.
Deploying $\Oc(10^4)$ tonnes of the cheaper element argon has already been proposed for the imminent DUNE experiment~\cite{DUNE:2020lwj,DUNEModuleDM:PNL2020,*DUNEModuleDM:snowmass:Avasthi2022,*DUNEModuleDM:Bezerra2023}.

As discussed in Refs.~\cite{DARWIN:Macolino:2020uqq,PandaX-xT:2024oxq,nufloor:GaspertGiampaMorrissey:2021gyj} and elsewhere, in the recoil energy ranges of interest the dominant irreducible NR background at next-generation detectors would be from coherent elastic recoils of atmospheric neutrinos.
Other backgrounds can be reduced, {\em e.g.}, NRs from radioactive decay neutrons via radiopure detectors and veto systems, and ERs from solar neutrinos and noble nuclide impurities via purification and ER rejection. 
Therefore we consider only atmospheric neutrino backgrounds and their uncertainties in our projections.
Using the neutrino fluxes and their conversion to the $\Erec$ spectrum in Ref.~\cite{nufloor:Billard:2013qya}, we obtain the values of $N_{\rm B}$ listed in Table~\ref{tab:detectordetails} at future experiments.
The NR acceptances $\epsilon_{\rm NR} = \{0.5, 0.5, 0.9, 0.9\}$ that we have used for \{DARWIN/XLZD, PANDAX-xT, DarkSide-20k, Argo\} assume ER rejections of $\{2 \times 10^{-4}, 3 \times 10^{-3}, 3\times10^{-8},  10^{-9}\}$~\cite{nufloor:GaspertGiampaMorrissey:2021gyj,PandaX-xT:2024oxq,DarkSide-20k:2017zyg}.
We will assume the  $\Erec$ range and NR acceptance vs ER rejection of DARWIN/XLZD and Argo for our hypothetical mega xenon and argon detectors as well.
The solid curves in Fig.~\ref{fig:nuroof} correspond to taking $N_{\rm obs} = N_{\rm B}$ in Eq.~\eqref{eq:def:90CL}, that is, by anticipating the number of observed events to be exactly the expected background count.
But the observed event count could fluctuate, therefore we also show, with dot-dashed curves on either side of the solid curve, the sensitivity for $N_{\rm obs} =  N_{\rm B}/2$ and  $N_{\rm obs} =  2N_{\rm B}$.
We expect the sensitivity that will be actually obtained to lie somewhere between these curves.
Masses reachable above the Planck scale would correspond to composite DM models~\cite{Models:Nuggets,Bramante:2018tos,Bramante:2019yss,snowmass:Carney:2022gse}. 

We see that increasing the exposure results in a xenon detector floor that is slightly lower and an argon detector floor that does not discernibly move.  
This is just what one expects deep within the neutrino fog, where the increase with exposure in neutrino backgrounds and the associated flux uncertainty limits the DM signal sensitivity; see Sec.~\ref{subsec:fog}.
We also see that, as discussed in Sec.~\ref{subsec:scalingtrends} and Sec.~\ref{subsec:limits},
larger detector masses (of the same material) set the single-scatter ceilings lower.
Another interesting effect is that higher exposures of the same detector material move the maximum mass to the left.
This is because of the increase in $N_{\rm exp}^{\rm 90CL}$ with the increasing background, so that $m_{\rm \chi}^{\rm max}$ in Eq.~\eqref{eq:mxturnaround} is reduced.
The trends of xenon vs argon sensitivities and comparison across the panels of Fig.~\ref{fig:limits} are the same as discussed in Sec.~\ref{subsec:limits}, and so we do not repeat them here.
The floors and ceilings shown can be trivially extrapolated to smaller DM masses using the linear and logarithmic scalings of Eqs.~\eqref{eq:sigTxeffvmdm-bot} and \eqref{eq:sigTxeffvmdm-top}; at any rate we show some ceilings down to TeV mass in Appendix~\ref{app:Mdetvariedroofs}.

We once again state the thought behind our showing the plots in Fig.~\ref{fig:nuroof}.
If detector exposures are increased, atmospheric neutrino backgrounds (as relevant for high DM mass) increase.
To claim a hint or discovery of DM, the number of signal events must be accordingly large as per  Poisson statistics (Eq.~\eqref{eq:def:90CL}).
If such an excess is found, whether or not the $(\sigmaNx, \mdm)$ producing it is near the single-scatter ceiling may be rapidly checked by the experiment by performing a multiscatter search {\em \`a la} Refs.~\cite{DEAP:MS:2021raj,XENON1T:MS:2023iku,LZ:MS:2024psa}.
From measurements of the integrated flux (Eq.~\eqref{eq:integfux}) and average multiplicity (Eq.~\eqref{eq:multiplicity}) one can even home in on the DM-nucleon cross section and mass, as detailed in Ref.~\cite{Bramante:2018qbc}. 
Therefore sensitivity plots such as in Fig.~\ref{fig:nuroof} would be the first step in verifying a DM signal at large cross sections and masses.
Of course, this sequence of steps holds for null results as well, which can be mutually cross-checked by single-scatter and multiscatter searches in overlapping regions of sensitivity.

\section{Discussion}
\label{sec:disc}

\subsection{Fog on the roof}
\label{subsec:fog}

In our work we have derived both upper and lower bounds on DM-nucleon scattering cross sections at existing and future noble liquid detectors.
Experiments from Gen-3 onward are expected to observe atmospheric neutrinos in coherent elastic recoils~\cite{nufloor:Newstead:2020fie}, forming a near-irreducible background to DM scatters, the ``neutrino floor''~\cite{nufloor:usage:CohenLisantiPierceSlatyer:2013}.
It has been known for some time that the neutrino floor is not a hard limit, which is best understood from the scaling of the DM sensitivity with the exposure or number of background events~\cite{nufloor:Billard:2013qya}.
For $N_{\rm BG} \ll 1$, this sensitivity scales as $\sigma_{\rm 90\% CL} \propto (\Mdet \texp)^{-1} \propto \NBG^{-1}$ as expected for a background-free search.
As exposures increase, the sensitivity is at first set by Gaussian fluctuations as $1/\sqrt{\Mdet \texp}$ before its improvement is impeded by background systematic uncertainties, so the scaling can be written as~\cite{nufloor:Billard:2013qya} 
\beq
\sigma_{\rm 90\% CL} \propto \frac{\sqrt{\NBG + \xi^2 \NBG^2}}{\NBG}~,
\eeq
where $\xi$ is the neutrino flux systematic uncertainty (here taken as 20\%).
We encountered these trends in Sec.~\ref{subsec:nuroofplots} when discussing the sensitivity floor.
If exposures further increase and $\gsim 10^4$ neutrino events are collected, there will be enough statistics to distinguish between the binned neutrino and DM recoil spectra, and a Gaussian scaling with exposure will be recovered. 
(This requires extremely optimistic exposures of $\gsim 10^6$ ton-yr, which we do not consider here.)
For these reasons the mapping from recoil spectra to the $\sigmaTx$-$\mdm$ plane has been recently dubbed the ``neutrino fog''~\cite{nufloor:fogusage:Akerib:2021pfd,nufloor:OHare:2021utq}.
Analogous considerations apply to the neutrino roof as well: it is more useful to think of a foggy ceiling than a hard roof where DM sensitivities weaken. 
However, the dependence of the sensitivity on $\NBG$, which in turn is related to $\NexpCL$ via Eq.~\eqref{eq:def:90CL}, is non-trivial due to the logarithmic scalings of the single-scatter ceiling cross section in Eq.~\eqref{eq:sigTxeffvmdm-top}, as also discussed in Sec.~\ref{subsec:nuroofplots}. 

\subsection{Raising the ceiling}
From Eq.~\eqref{eq:sigTxeffvmdm-top} we can envisage a few ways to push the single-scatter ceiling up, thereby probing more real estate in DM cross section vs mass space.
In already existing data the ``region of interest'' (ROI) of single-scatter (WIMP) searches can be redefined in two ways.

[i] Use a volume of the detector smaller than the fiducial one, effectively reducing $L_{\rm ave}$.
The search volume may be shrunk to a size comparable to the spatial resolution to which the interaction vertex may be reconstructed.
This is $3 \times 3 \times 1$~mm$^3$ for xenon~\cite{XENONxyzreconstruct:2006gfg} and $6 \times 6 \times 1$~mm$^3$ for argon time projection chambers (TPCs)~\cite{DarkSide:SS:2018kuk} (where the Ar $z$-position resolution is obtained from the $\mu$s binning of the electron drift time in Ref.~\cite{DarkSidexyreconstruct:Watson:2017soo} and the drift speed 1~mm/$\mu$s), implying that the ceiling in Eq.~\eqref{eq:sigTxeffvmdm-top} may be raised by orders of magnitude. 
As the detector effective area is also shrunk, this would come at the cost of sensitivity to large DM masses (Eq.~\eqref{eq:mxturnaround}).

[ii] Narrow the range of recoil energies down to the energy resolution, which is a few keV for xenon and argon in the $\Erec$ ranges considered here~\cite{nufloor:GaspertGiampaMorrissey:2021gyj}.
This would raise the single-scatter ceilings by $O(1)$ factors without costing sensitivity to large DM masses.

In principle, limits from multiple of these small ROIs may be combined to improve the net sensitivity.
We do not attempt to push our ceilings up by recasting searches using [i] and [ii] as it is unclear which of the redefined ROIs observed events in the datasets fall into. 
As such, the single-scatter ceilings we have drawn, as well as those displayed by Refs.~\cite{XENON1T:MS:2023iku,LZ:MS:2024psa}, are conservative limits.

Eq.~\eqref{eq:sigTxeffvmdm-top} implies that per-nuclear cross section ceilings can also be raised by considering detectors with less dense material, {\em e.g.}, gaseous detectors.
If realized, kilotonne gaseous xenon TPCs~\cite{Xekton:Avasthi:2021lgy,*Xekton:Anker:2024xfz} would be profitable in this regard.
CYGNUS~\cite{CYGNUS:2020pzb}, made of mostly helium with number density $4 \times 10^{19}~{\rm cm}^{-3}$ and with $L_{\rm ave} \simeq 10$~m, would give ceilings $\Oc(10^2)$ times higher than liquid xenon detectors with target number density = $10^{22}~{\rm cm}^{-3}$ and $L_{\rm ave} = \Oc(1)$~m.
For $A^4$ scaling scenarios, the per-nucleon ceiling would be a further $(132/4)^4 \simeq 10^6$ times higher.
In these cases, due to the large detector volume, the maximum reachable DM mass would be comparable to that of next-generation noble liquid detectors.
To a lesser yet notable extent, NEWS-G~\cite{NEWS-G:2022kon}, a spherical proportional counter (SPC) made of mostly neon with number density $4 \times 10^{19}~{\rm cm}^{-3}$ and with diameter 130~cm, would also break the gas ceiling.

Attempts to improve DM sensitivity below the neutrino floor using different target nuclei, annual modulation, neutrino directionality and refinements in neutrino flux measurements~\cite{nufloor:Billard:2013qya,nufloor:OHare:2020lva,nufloor:Snowmass:Akerib2022} would apply to the neutrino roof as well, but to a lesser extent due to the logarithmic scalings of the ceiling with event counts as in Eq.~\eqref{eq:sigTxeffvmdm-top}.

\begin{figure*}[th]
    \centering
    \includegraphics[width=.47\textwidth]{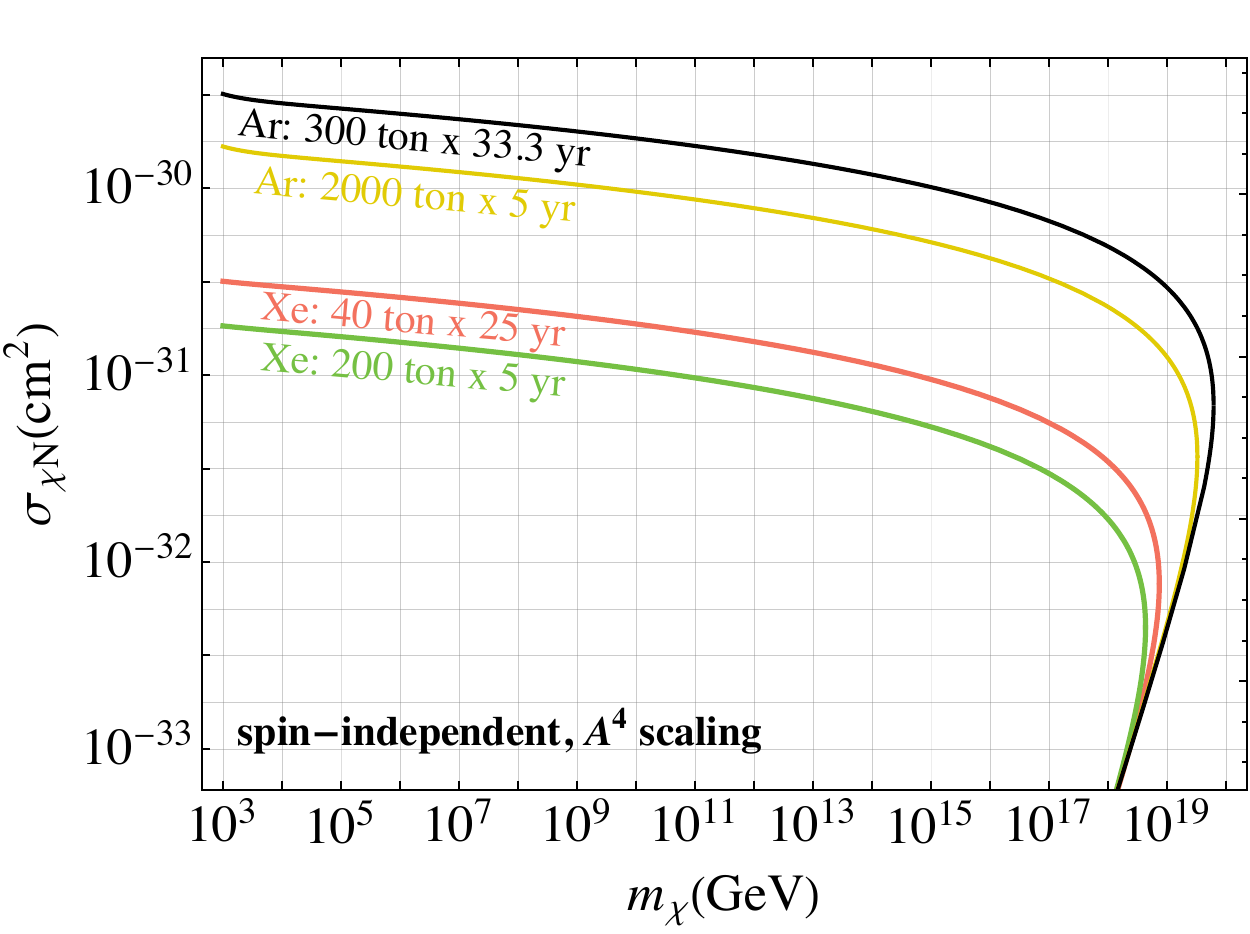}  \includegraphics[width=.47\textwidth]{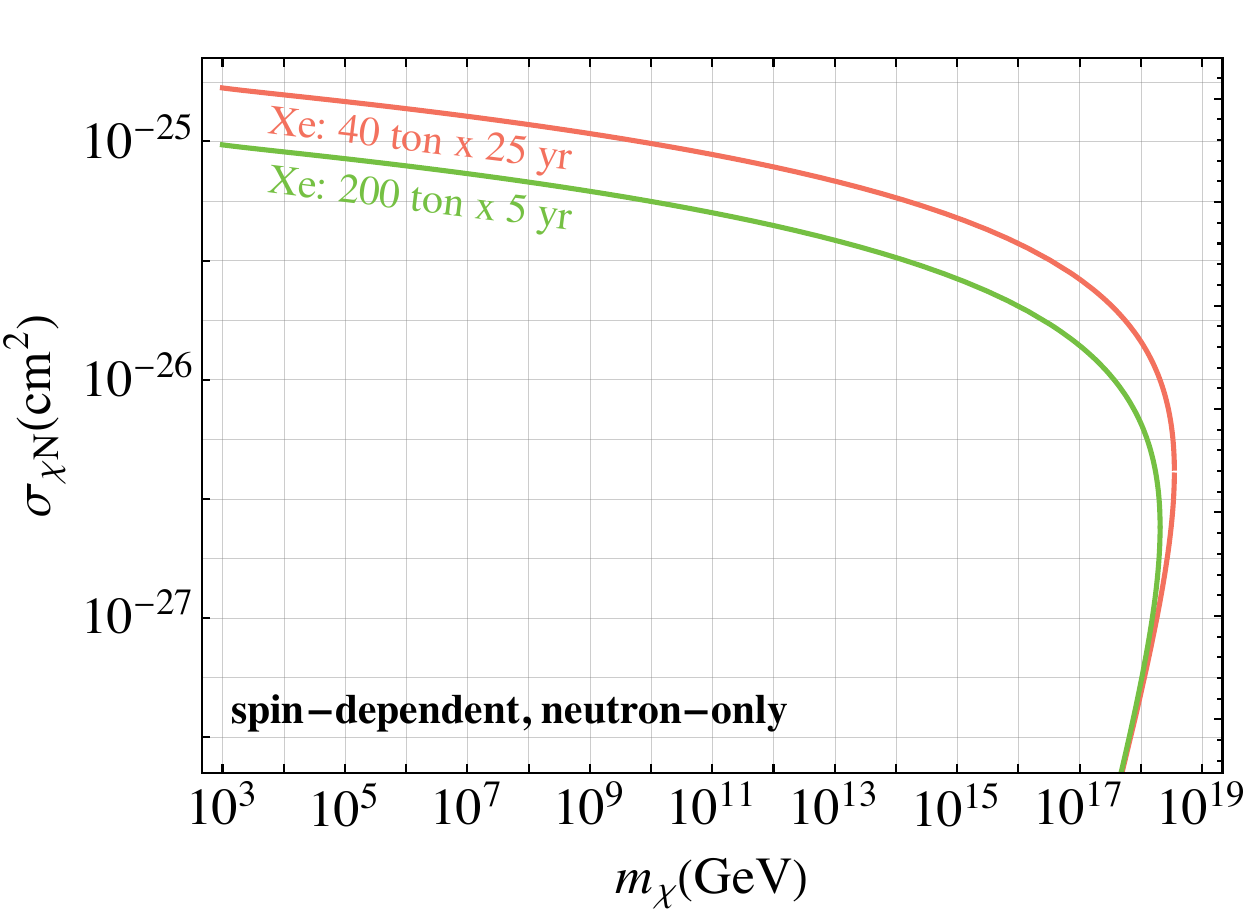}
    \caption{Illustrations that single-scatter ceilings exhibit the scalings in Eqs.~\eqref{eq:sigTxeffvMdettexp} for different detector configurations $(\Mdet, \texp)$ of the same material with fixed exposure $\Mdet \times \texp$. 
    At small cross sections the floors (upper bounds) do not show a difference in sensitivity, which simply scale as $(\Mdet \texp)^{-1}$. 
     Much of the parameter space shown here is ruled out by earlier searches~\cite{KavanaghSuperheavy:2017cru,gascloudcompendium:Bhoonah:2018gjb,PlasticEtch:Bhoonah2020fys,snowmass:Carney:2022gse}.}
    \label{fig:nuroofMTvariations}
\end{figure*}

\subsection{Future scope}

Considerations of single-scatter ceilings open up many paths of research.
Whilst we have explored these ceilings at high DM masses accounting for atmospheric neutrinos, deriving them at low DM masses accounting for the flux of less energetic solar and diffuse supernova neutrinos would be an interesting and non-trivial exercise.
At the same time, detectors of every kind could be brought under one (neutrino) roof: 
the current 
noble liquid,
(scintillating) bubble chamber,
solid state and 
SPC technologies,
and the future
snowball chambers, 
giant gas TPCs, 
solid xenon,
archaeological lead,
DUNE-like modules,
novel condensed matter systems, 
and so on;
see Refs.~\cite{nufloor:Snowmass:Akerib2022,SBC:snowmass2022,bubblechambersSnowmassLOI,*superheatedliquidSethDas2019, lowthresh:snowmassEssig2022,snowmass:Carney:2022gse, snowmassreportpclDMCooley2022,*Xekton:Anker:2024xfz,RESNOVA:Pattavina:2020cqc,DUNEModuleDM:PNL2020,*DUNEModuleDM:snowmass:Avasthi2022,*DUNEModuleDM:Bezerra2023,lightweightZurek2024} and references therein.
Paleo-detectors, shown to be sensitive to variations in atmospheric neutrino fluxes over Gyr timescales~\cite{paleo:Jordan:2020gxx} and DM scattering~\cite{paleo:Baum:2024eyr}, would also exhibit DM single-scatter ceilings, which may be derived using our formalism and displayed in future studies.
More futuristically, direct detection on the Moon would encounter a background of neutrinos made by cosmic ray interactions with the regolith~\cite{nufloor:Moon:TRIUMF2023}, and a neutrino roof may be estimated for such experiments.
In the cases above, single-scatter ceilings would seem to largely overlap with parameter space excluded by earlier experiments,
but as we have reiterated throughout this work, it is important to identify the full extent of the constraints for DM scattering on individual nuclear species.

Single-scatter ceilings may also be estimated for DM-electron scattering, and thence a neutrino roof for electron recoils may be identified in the style of an analogous neutrino floor/fog~\cite{nufloor:electron:Carew:2023qrj}.
Our formalism can be extended to inelastic DM~\cite{inelastic:Tucker-Smith:2001myb,*inelastic:Bramante:2016rdh}.
The elastic scatters we have considered would be the limit of small mass splittings between the ground and excited states of DM; in the optically thick limit there will be an alternating succession of endothermic and exothermic scatters of these states, which may give rise to interesting multiscatter signals that may cross-check results near single-scatter ceilings.
Another model assumption that has gone into our treatment is a velocity-independent contact interaction between DM and nucleons that gives rise to the differential cross section in Eq.~\eqref{eq:opticallythinlimit}.
But velocity-dependent and/or light-mediator interactions would alter $d\sigmaTx/d\Erec$.
Similarly, the form factor would be modified for DM-pion scattering~\cite{XENONDMpion:2018clg} and certain composite DM models~\cite{Models:nucleiHardy:2014mqa,*Models:nucleiHardy:2015boa,*Models:nucleiMonroe:2016hic,*Models:Nuggets}.

Finally, these cases may be treated for a cosmological relic that make up a fraction $f_{\rm DM}$ of the DM density.
The single-scatter ceilings would scale as $\log f_{\rm DM}$ (Eq.~\eqref{eq:sigTxeffvmdm-top}) while the floors lose sensitivity with the expected scaling of $f_{\rm DM}^{-1}$ (Eq.~\eqref{eq:sigTxeffvmdm-bot}).

\subsection{Concluding remarks for experimental collaborations}

We strongly urge dark matter direct detection experiments to perform the following three steps after every data-taking run.
The first two require negligible effort but would present the maximum impact of their experimental set-up and WIMP search data analysis.
The third requires dedicated analysis and would expand the discovery reach of their detector by several orders of magnitude.
These points have been also been made before~\cite{snowmass:Carney:2022gse}, but they bear repeating.

(i) Extend WIMP reaches to DM masses higher than the usual $10^3-10^4$~GeV presented, and go all the way to the highest DM mass reachable (Eq.~\eqref{eq:mxturnaround}), which would be typically $> 10^{16}$~GeV. 
Note: the unitarity bound $\mdm \leq \Oc(100)$~TeV does not apply to the non-thermal DM candidates listed in the Introduction.

(ii) Estimate and display the single-scatter ceiling using the prescription in this work. Accounting for detector geometry as in Refs.~\cite{XENON1T:MSSSprojexn:Clark:2020mna,XENON1T:MS:2023iku,LZ:MS:2024psa} would refine this bound.

(iii) Perform a multiscatter search in the high-mass region; noble liquid and bubble chamber collaborations may build on the methods described in Appendix~\ref{app:searchmethods} and Ref.~\cite{PICO:MS:Broerman2022}. Excluded regions from single-scatter and multiscatter searches that overlap would mutually confirm null results.
Signal hints in single-scatter searches may be confirmed in multiscatter searches, and the DM-target cross section and DM mass may be pinpointed as outlined at the end of Sec.~\ref{subsec:nuroofplots} and in Ref.~\cite{Bramante:2018qbc}.
\vspace{.3cm}

We look forward to experiments pursuing dark matter all the way up to the neutrino roof.

\section*{Acknowledgments}

Many thanks to 
Sayan Ghosh 
and
Ibles Olcina Samblas
for insightful discussions.

\appendix

\section{Dependence on detector mass and live time}
\label{app:Mdetvariedroofs}

See Fig.~\ref{fig:nuroofMTvariations} for an illustration of the variation of the single-scatter ceiling with detector configurations giving the same exposure, as mentioned in Sec.~\ref{subsec:nuroofplots}.

\section{Search methods}
\label{app:searchmethods}

We briefly review here the search techniques of noble liquid-based direct detection.
Detailed descriptions can be found in the references in Table~\ref{tab:detectordetails} and Refs.~\cite{APPEC:Billard:2021uyg,wpXENON:Aalbers:2022dzr}.

Xenon detectors are dual phase TPCs with arrays of photomultiplier tubes (PMTs) at the top and bottom.
Recoils produce prompt scintillation (S1) in the liquid phase and proportional scintillation (S2) in the gaseous phase on top.
In single-scatter searches, NRs such as those induced by DM scattering are discriminated from ER backgrounds using the ratio S2/S1, with the former giving lower values.
In XENON1T's multiscatter search~\cite{XENON1T:MS:2023iku}, which partially follows the strategy outlined in Ref.~\cite{Bramante:2018qbc}, signals were marked mainly in terms of S2: large total area and total width of S2 pulses, a higher ratio of top vs bottom PMTs being lit up by S2 than by S1, and a wider pulse for S2 ($\sim 10~\mu$s) than for possibly merged S1s (0.2$-$0.8~$\mu$s).
In LZ's multiscatter search~\cite{LZ:MS:2024psa} deliberately lenient cuts were chosen, resulting in the visibly stronger bounds than XENON1T in Fig.~\ref{fig:limits} at low cross sections.
Signals were identified in terms of the number of S1s and S2s, and the track's collinearity, uniformity of distribution of scatters, and velocity.

The argon detector DarkSide-50 is a dual phase TPC, with its single-scatter search~\cite{DarkSide:SS:2018kuk} performed using the NR/ER pulse shape discrimination (PSD) variable $f_{90}$, the S1 light collected in the first 90~ns of an event, which is higher for NRs.
DarkSide-20k will use the analogously defined $f_{200}$~\cite{DarkSide-20k:2017zyg}.
DEAP-3600 is a liquid argon detector that uses the PSD variable $F_{\rm prompt}$ which is the ratio of light collected in the intervals [-28~ns, 150~ns] and [-28 ns, 10$^4$~ns] of an event.
Its single-scatter search~\cite{DEAP:SS:2019yzn} identified NR signals as $F_{\rm prompt} \sim 0.7$ and photoelectron count (PE) $\lsim 200$, and ER backgrounds as $F_{\rm prompt} \sim 0.3$ and PE $\gsim 300$.
Its multiscatter search~\cite{DEAP:MS:2021raj} identified signals in the PE range $4\times10^3-4\times10^8$ with {\em small} $F_{\rm prompt}$ ($\leq 0.1$ for the most part) and well-separated peaks in the waveform.
PSD in argon gives an ER rejection rate 5$-$6 orders of magnitude greater than S2/S1 in xenon as seen in Sec.~\ref{subsec:nuroofplots}, which would make Argo single-scatter sensitivities comparable to DARWIN/XLZD and PANDAX-xT.

\bibliography{refs}

\end{document}

%% file: universalnewcommands.tex
\newcommand{\gsim}{\gtrsim}
\newcommand{\lsim}{\lesssim}

\def\Oc{\mathcal{O}}

\newcommand{\beq}{\begin{equation}}
\newcommand{\eeq}{\end{equation}}
\newcommand{\bea}{\begin{eqnarray}}
\newcommand{\eea}{\end{eqnarray}}
\newcommand{\nn}{\nonumber}

\definecolor{rosy}{RGB}{230,235,252}
\definecolor{myframetitle}{RGB}{90,89,170}
\definecolor{myblocktitle}{RGB}{140,185,249}
\definecolor{mytitle}{RGB}{10,80,26}

\definecolor{darkgreen}{RGB}{27,130,45}
\definecolor{darkblue}{rgb}{0,0,0.3}
\definecolor{darkred}{rgb}{0.7,0,0}

\definecolor{light gray}{RGB}{220,220,220}
\definecolor{dark purple}{RGB}{108,0,217}
\definecolor{pink}{RGB}{190,20,100}
\definecolor{orang}{RGB}{193,63,0}
\definecolor{green}{RGB}{11,98,17}
\definecolor{darkpink}{RGB}{153,0,76}
\definecolor{bluegreen}{RGB}{0,102,102}
\definecolor{greenlagan}{RGB}{0,102,0}
\definecolor{redgreen}{RGB}{102,102,0}
\definecolor{Redgreen}{RGB}{153,76,0}
\definecolor{vividviolet}{rgb}{0.62, 0.0, 1.0}
\definecolor{amaranth}{rgb}{0.9, 0.17, 0.31}
\definecolor{palatinateblue}{rgb}{0.15, 0.23, 0.89}
\definecolor{brightpink}{rgb}{1.0, 0.0, 0.5}
\definecolor{cornflowerblue}{rgb}{0.39, 0.58, 0.93}
\definecolor{deepcarminepink}{rgb}{0.94, 0.19, 0.22}
\definecolor{radicalred}{rgb}{1.0, 0.21, 0.37}